\documentclass[showpacs,preprintnumbers,amsmath,amssymb]{revtex4}

\usepackage{graphicx}
\usepackage{dcolumn}
\usepackage{bm}

\newcommand{\be}{\begin{equation}}
\newcommand{\ee}{\end{equation}}

\begin{document}

 
\title{\bf LATTICE BOLTZMANN MODELS FOR NON-IDEAL FLUIDS WITH ARRESTED PHASE-SEPARATION}

\author{S. Chibbaro}
\affiliation{Dept. of Mechanical and Industrial Engineering, University of ``Tor Vergata'', via del politecnico 1 00133, Rome, Italy}
\author{G. Falcucci}
\affiliation{Dept. of Mechanical and Industrial Engineering, University of ``Roma Tre'', Via della Vasca Navale 79 00146 Rome, Italy}
\author{H. Chen}
\affiliation{EXA Corporation, 3 Burlington Woods Drive, Burlington, MA 01803, USA}
\author{X. Shan}
\affiliation{EXA Corporation, 3 Burlington Woods Drive, Burlington, MA 01803, USA}
\author{G. Chiatti}
\affiliation{Dept. of Mechanical and Industrial Engineering, University of ``Roma Tre'', Via della Vasca Navale 79 00146 Rome, Italy}
\author{S. Succi}
\affiliation{Istituto Applicazioni Calcolo, CNR, V.le del Policlinico 137,
00161, Rome, Italy}
\date{\today}

\begin{abstract}
The effects of mid-range repulsion in Lattice Boltzmann models on the 
coalescence/breakup behaviour of single-component, non-ideal fluids are investigated. 
It is found that mid-range repulsive interactions allow the formation of  
spray-like, multi-droplet configurations, with droplet size directly related 
to the strength of the repulsive interaction. 
The simulations show that just a tiny ten-percent of mid-range repulsive pseudo-energy 
can boost the surface/volume ratio of the phase-separated fluid 
by nearly two orders of 
magnitude.
Drawing upon a formal analogy with magnetic Ising systems, a 
pseudo-potential energy is defined, which
is found to behave like a quasi-conserved quantity for most 
of the time-evolution.
This offers a useful quantitative indicator of the stability
of the various configurations, thus helping the task
of their interpretation and classification. 
The present approach appears to be a promising tool for the 
computational modelling of complex flow phenomena, such as 
atomization, spray formation and micro-emulsions, break-up phenomena 
and possibly glassy-like systems as well.

\end{abstract}


\maketitle


\section{Introduction}

In the last two decades, the
Lattice-Boltzmann (LB) approach has emerged as powerful mesoscopic alternative 
to classical macroscopic methods for computational hydrodynamics \cite{MZ,HSB,Succi_book}.
The pseudopotential method, put forward a decade ago by X. Shan and H. Chen to endow Lattice Boltzmann
models with potential energy interactions, is one of the most successful outgrowths of basic LB theory \cite{SC_93, SC_94}.
The Shan-Chen (SC) model is based on the idea of representing intermolecular
interactions at the mesoscopic scale via a density-dependent nearest-neighbour
pseudopotential $\psi(\rho)$.
Despite its simplified character, the SC model provides the 
essential ingredients
of non-ideal fluid behaviour, namely a non-ideal equation of state and surface
tension effects at phase interfaces.
Due to its remarkable computational simplicity, the SC method is 
being used for a wide and 
growing body of complex flows applications, such as multiphase flows in chemical, manufacturing and geophysical problems.

To date, the overwhelming majority of Shan-Chen applications have been
performed within the original formulation, whereby only 
first-neighbor attractive interactions are included. 
This entails a number of limitations, primarily
the impossibility to tune the surface tension independently of the
equation of state. This limitation has been recently lifted by introducing
second-neighbor {\it repulsive} interactions \cite{Fal_07}.
Besides offering an independent handle on the surface tension, it has 
been observed that second-neighbor (mid-range) repulsion may disclose
an entirely new set of physical regimes, particularly the onset
of metastable multi-droplet configurations, which would be impossible
to obtain with short-range attraction alone. 
These configurations result from the existence of energy barriers
(mid-range repulsion) which slow-down/arrest the dynamics of coarsening/phase-separation \cite{seth1,seth2}
In this work, we provide a quantitative exploration of the basic 
mechanisms behind this physically enriched scenario.
To this aim, we investigate the structural properties of multi-droplet
configurations, as well as their energetics, as a function of the
main parameters of the model, mainly the strength of the 
repulsive interactions.
Upon progressive switching of this paramater, the system is found 
to move from a single-droplet phase-separated fluid, to a multi-droplet
metastable configurations, all the way up to a quasi-ordered
crystal-like structure. 

\section{Standard Shan-Chen model}

The standard lattice Boltzmann (LB) equation with pseudopotential interaction can be expressed as follows,
\be
\label{eq.1}
f_i(\vec{x}+\vec{c}_i,t+1)-f_i(\vec{x},t) = -\omega(f_i-f_i^{eq}) + {F_i}(\vec{x})
\ee
where $f_i$ is the probability density function of finding a particle at site $\vec{r}$ at time $t$, moving along the $i-th$ lattice direction defined by the discrete speeds $\vec{c}_i$ with $i=0,...,b$. The left hand-side of (\ref{eq.1}) stands 
for molecular free-streaming, whereas the right-hand side represents the time relaxation (due to collisions) towards local Maxwellian equilibrium. Finally, $F_i$ represents the total volumetric body force. In particular, we shall use a {\it dynamic mean-field} term connected with bulk particle-particle interactions.
The macroscopic density $\rho$ and velocity $\vec{u}$ are given by \cite{Succi_book}

\be
  \rho(\vec{x},t) = \sum_{i=0}^b f_i
\ee
\be
  \rho(\vec{x},t)\vec{u}(\vec{x},t) =   \sum_{i=0}^b \vec{c}_i f_i
\ee

The equilibrium distribution function is calculated in order to make the collision operator conserve mass and momentum: a common choice that satisfies the above constraints is the following

\be
  f_i^{eq}=w_i \rho \bigg(1 + \frac{1}{c_s^2} \vec{c}_i \cdot \vec{u} + \frac{1}{2 c_s^4} (\vec{c}_i \cdot \vec{u})^2 -
  \frac{1}{2 c_s^2} u^2 \bigg)
\ee
The $F_i$ term in (\ref{eq.1}) represents the phase interaction, 
\be 
  \vec F = - G \psi(\vec{x}) \sum_{i=0}^b w_i \psi(\vec{x}+\vec{c}_i)\vec{c}_i
\label{forza}
\ee
in which $\psi(\rho)$ is the local pseudopotential governing the interaction
and  $w_i$ are statistical weights which will be defined in the following.
The expression of $\psi(\rho)$ by Shan and Chen is the following,

\be
  \psi=\sqrt{\rho_0} (1-e^{-\rho/\rho_0})
\ee
In this model, phase separation is achieved by imposing a short-range attraction between the
light and dense phases. Indeed, such short-range attraction  is responsible for
the growth of density contrasts through a dynamical instability of the
interface. 
In real fluids, such instability is  tamed by hard-core
repulsion, while in the SC model such hard-core repulsion is not included,
for it would impose significant penalty on the time-marching procedure, 
and is replaced instead by a saturation
of the attractive interactions above a given density threshold, $\rho_0$. 
Expanding $F_i$, (\ref{forza}), in terms of $\vec{c_i}$, we find, to fourth  order \cite{He98}

\be
\label{eq:fsc}
  \vec F= -C_1 c_s ^2 G  \psi  \vec{\nabla}\psi -C_2 c_s^4 G  \psi  \vec{\nabla}\nabla^2\psi
\ee
where $C_1$ and $C_2$ are lattice-specific numerical factors. The first term is responsible for the non-ideal part of the corresponding
 equation of state:

\be
\label{pressure}
 P = \rho  c_s^2 +  \frac{1}{2}  C_1 c_s^2 G  \rho \ \psi^2~.
\ee
The second term in eq. (\ref{eq:fsc}) is the inherent surface tension in the SC model which yields
\be
\gamma = - \frac{C_2 G}{2} c_s^4 \int_{-\infty}^{\infty} |\partial_y \psi)|^2 dy
\label{eq:gammasc}
\ee
Considering  the interface-equilibrium condition, $ \frac{1}{c_s^2} \frac{\partial p}{\partial \rho} = \frac{1}{c_s^2} \frac{\partial^2 p}{\partial \rho^2}= 0$, we find the critical condition for phase separation, 
$G<G_{cr} = -4.0$, $\rho_{cr} = \rho_0 \ ln2$

\section{Short and mid-range interactions}

Our model is based on the interaction between each particle and a 
set of $24$ surrounding neighbours, 
distributed over two Brillouin zones (\textit{belts} for simplicity).
The interaction force in (\ref{eq.1}) reads as follows 
\be
\label{PFORCE}
   \vec{F}(\vec{x}) = \sum_{k=s,m}  G_k \psi_k(\rho(\vec{x})) \sum_{i=1}^{b_k} p_{ki} \vec{c}_{ki}
\psi_k(\rho(\vec{x}+\vec{c}_{ki}))
\ee
where the index $k=s,m$ labels the short and mid range belts respectively, whereas 
$\vec{c_{ki}}$ denotes the $i$-th set of discrete speeds belonging to the $k$-th belt.
The pseudo-potential force consists of two separate components
$\vec{F}(\vec{x}, t) = \vec{F}_s(\vec{x}, t) + \vec{F}_m(\vec{x}, t)$, defined as follows:  
\begin{eqnarray}
\label{FORCE}
\vec{F}_s(\vec{x}, t) &=& G_1 \psi(\vec{x},t) \sum_{i=1}^{b_1} w_i \psi(\vec{x}_{i},t) \vec{c}_{i} \Delta t \\
 &+& G_2 \psi(\vec{x},t) \sum_{i=1}^{b_1} p_{si} \psi(\vec{x}_{si},t) \vec{c}_{si} \Delta t \nonumber \\ 
\vec{F}_m(\vec{x}, t) &=& 
  G_2 \psi(\vec{x},t) \sum_{i=1}^{b_2} p_{mi} \psi(\vec{x}_{mi},t) \vec{c}_{mi} \Delta t \nonumber
\end{eqnarray}
In the above, $w_i$ are the weights of the first belt of neighbours, the same as in the standard SC model; the indices $k=s,m$ refer to the first and second Brillouin belts in the lattice, 
and $\vec{c}_{ki}$, $p_{ki}$ are the corresponding discrete speeds and 
associated weights, reported in Tab. 1.
Finally $\vec{x}_{ki} \equiv \vec{x}+ \ \vec{c}_{ki} \Delta t$ are the displacements
along the $i$-th direction in the $k$-th belt.

\begin{figure}
 \includegraphics[scale=0.4]{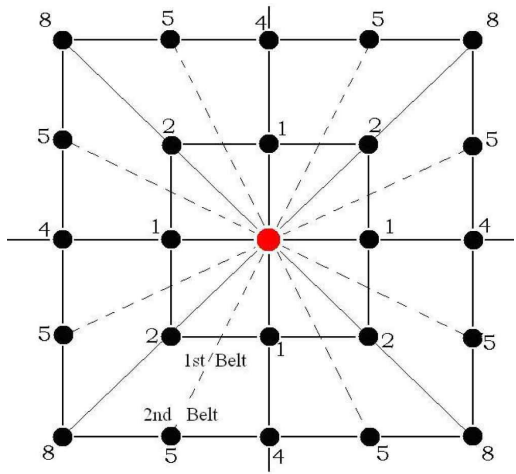}
\caption{\small {(Color Online) Two-belt lattice for force evaluation. Each node is labelled by the
 corresponding energy $|c_{ki}|^2$. Belt $1$ contains eight speeds and two energy levels $(1,2)$. 
Belt $2$ contains sixteen speeds, distributed over three energy levels $(4,5,8)$}}
\label{Fig:2-belt}
\end{figure}

\begin{table}[!h]
\begin{center}
\begin{tabular}{l l}
\hline
 \qquad \qquad \qquad E(8) \\
\hline
  $p_{si}  = p(1) = 4/63\; , $&$\quad i = 1,4$\\
  $p_{si}  = p(2) = 4/135 \; , $&$\quad i = 5,8$\\
  $p_{mi} = p(4) = 1/180 \; , $&$\quad i = 1,4$\\
  $p_{mi} = p(5) = 2/945 \; , $&$\quad i=5,12$\\
  $p_{mi} = p(8) = 1/15120 \; , $&$\quad i = 13,16$\\
\hline 
\end{tabular} 
\end{center}
\caption{\small{Links and weights of the two-belt, 24-speed lattice\cite{Shan_06,Sbr_07}.}}
\end{table}

Note that $G$ is a measure of potential to thermal energy ratio, and positive(negative) $G$ 
correspond to repulsion(attraction) respectively.
The first belt is discretized with $9$ speeds ($b_1=8$), while the second with $16$ ($b_2=16$) 
and the weights are chosen in such a way as to fulfill the following normalizations \cite{Shan_06,Sbr_07}: 
\be
\label{weights_1}
 \sum_{i=0}^{b_1} w_i = \sum_{i=0}^{b_1} p_{si} + \sum_{i=1}^{b_2} p_{mi} = 1
\ee
\be
\label{weights_2}
  \sum_{i=1}^{b_1} w_i c_i^2 = \sum_{i=1}^{b_1} p_{si} c_{si}^2 + \sum_{i=1}^{b_2} p_{mi} c_{mi}^2= c_s^2
\ee
where $c_s^2=1/3$ is the lattice sound speed.
Note that the present set of discrete speeds and weights secures \textit{$8-th$} order isotropy in the force evaluation.
The pseudo-potential $\psi(\vec{x}) $ is taken in the form first suggested by Shan and Chen \cite{SC_93},
$\psi[\rho] = \sqrt{\rho_0} (1-e^{-\rho/\rho_0})$ where $\rho_0$ marks the density value (\textit{critical})
at which non ideal-effects come into play and it is fixed to $\rho_o=1$ in lattice units.
Taylor expansion of (\ref{FORCE}) to second-order delivers the following  
non-ideal equation of state (EOS) 

\be
 p \equiv P/c_s^2 = \rho + \frac{(g_1+g_2)}{2} \psi^2(\vec{x},t)
\ee
where $g_k \equiv G_k /c_s^2$ are normalized coupling strengths.
Further expansion of eq. (\ref{FORCE}) to fourth-order provides the following expression for the surface tension
\be
\gamma = - \frac{(G_1+ \frac{12}{7} G_2)}{2} c_s^4 \int_{-\infty}^{\infty} |\partial_y \psi)|^2 dy
\label{eq:gamma}
\ee
where $y$ runs across the phase interface.
This is the analogue of eq. (\ref{eq:gammasc}), with the correspondence : $C_1G \leftrightarrow (G_1+G_2)$ and $C_2G \leftrightarrow (G_1+\frac{12}{7} G_2)$.

\section{Numerical results}

With two parameters at our disposal, $G_1$ and $G_2$, the present model allows 
a separate control of the equation of state and surface tension, respectively.
In particular, as shown in previous work \cite{Chi_07}, the non-ideal part
of the equation of state depends only on $A_1 = G_1 + G_2$,
whereas surface tension effects are controlled by the combination $ G_1 + \frac{12}{7} G_2$. 
Since in the vicinity of $\gamma \rightarrow 0$  higher order terms come into play, it proves expedient to define a new coefficient 
\be
A_2=G_1+\lambda G_2
\ee
where the numerical factor $\lambda$ plays the role of a renormalisation 
parameter, whose departure from zeroth-order value $\frac{12}{7}$
is a measure of the influence of the higher-order terms.
Comparison with numerical results shows that  $\lambda\approx3/2$ provides satisfactory agreement, see eq. (\ref{eq:l}).
This shows that, at a given value of $A_1$ (i.e. given density ratio between the light
and dense phase), mid-range repulsion ($G_2>0$) is expected to lower the surface
tension of the fluid, thereby facilitating the formation of multi-droplet configurations
with higher surface/volume ratioes than the standard Shan-Chen model. 
Thus, the mid-range potential is expected to act as a ``surfactant'' \cite{degennes, colloids, sciortino}, where ``surfactant'' indicate that true surfactants can be transported by many different mechanisms, locally changing the surface tension of the fluid, which is not what the mid-range repulsion in the present work does.
Numerically, the role of the mid-range is to add higher-order derivative to standard interaction force, which provides more isotropy and enables control of the equilibrium surface tension. 
In order to explore this scenario, we have simulated
droplet formation by integrating the LBE Eq.(\ref{eq.1}) in a 2D lattice using the nine-speed 2DQ9 model 
\cite{Ben_92,Chen_98,Wolf},  out of a noisy density background ($\delta{\rho}/\rho \sim 0.01$) with initial
density $\rho_{in}=\rho_{0} ln2 + \delta \rho$ in a periodic domain. 
In all simulations, $\tau=1$.
We have performed a systematic scan over the force strength, by changing $G_1$ and
$G_2$ so as to keep $A_1 = -4.9$ while increasing $A_2$ above the Shan-Chen value $A_2=A_1=-4.9$. 
All simulations have been performed with a resolution of $512^2$ grid points,
and a total simulation time $t=500 000$.

\begin{figure}
(a)
\includegraphics[scale=0.28]{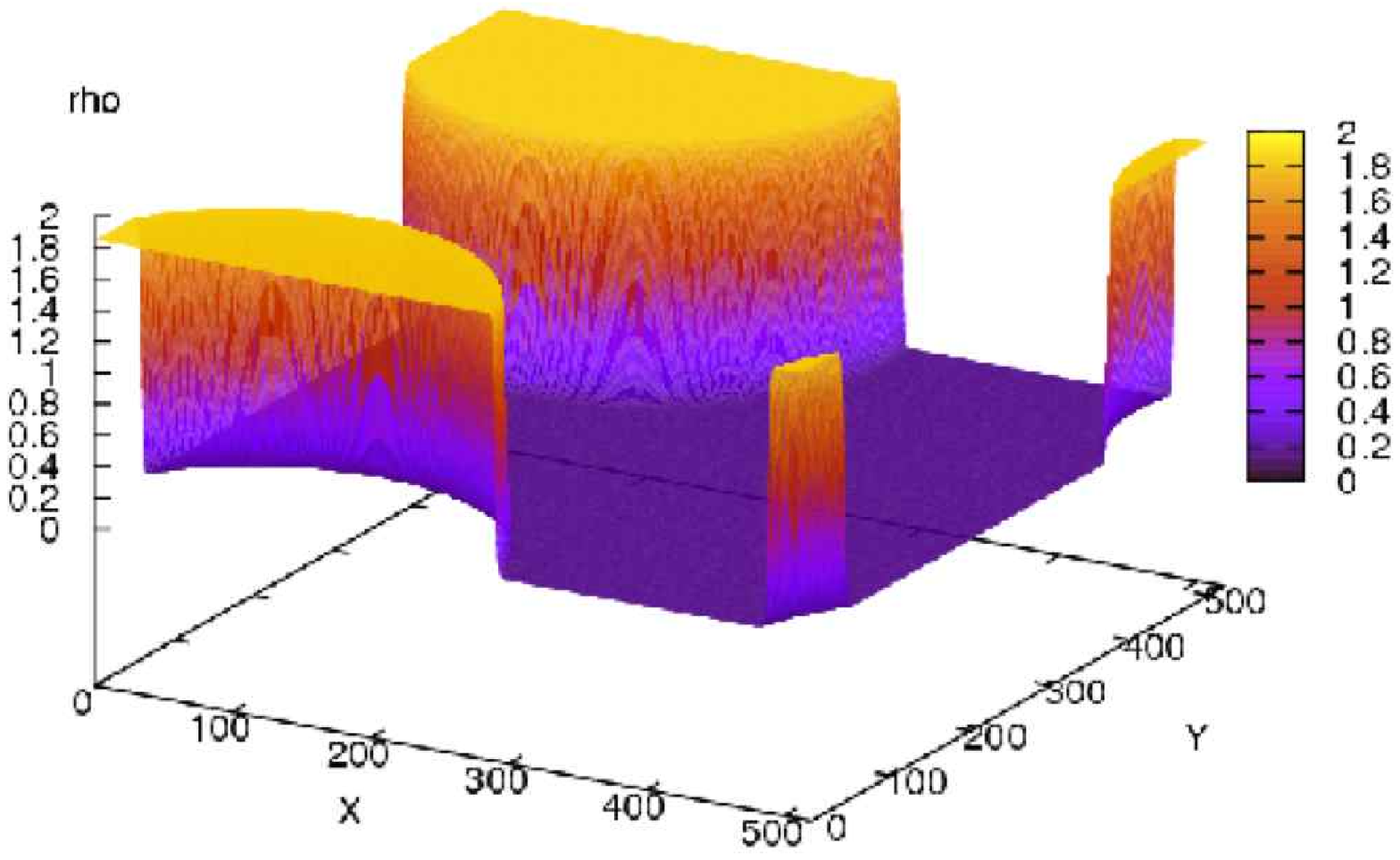}
\includegraphics[scale=0.28]{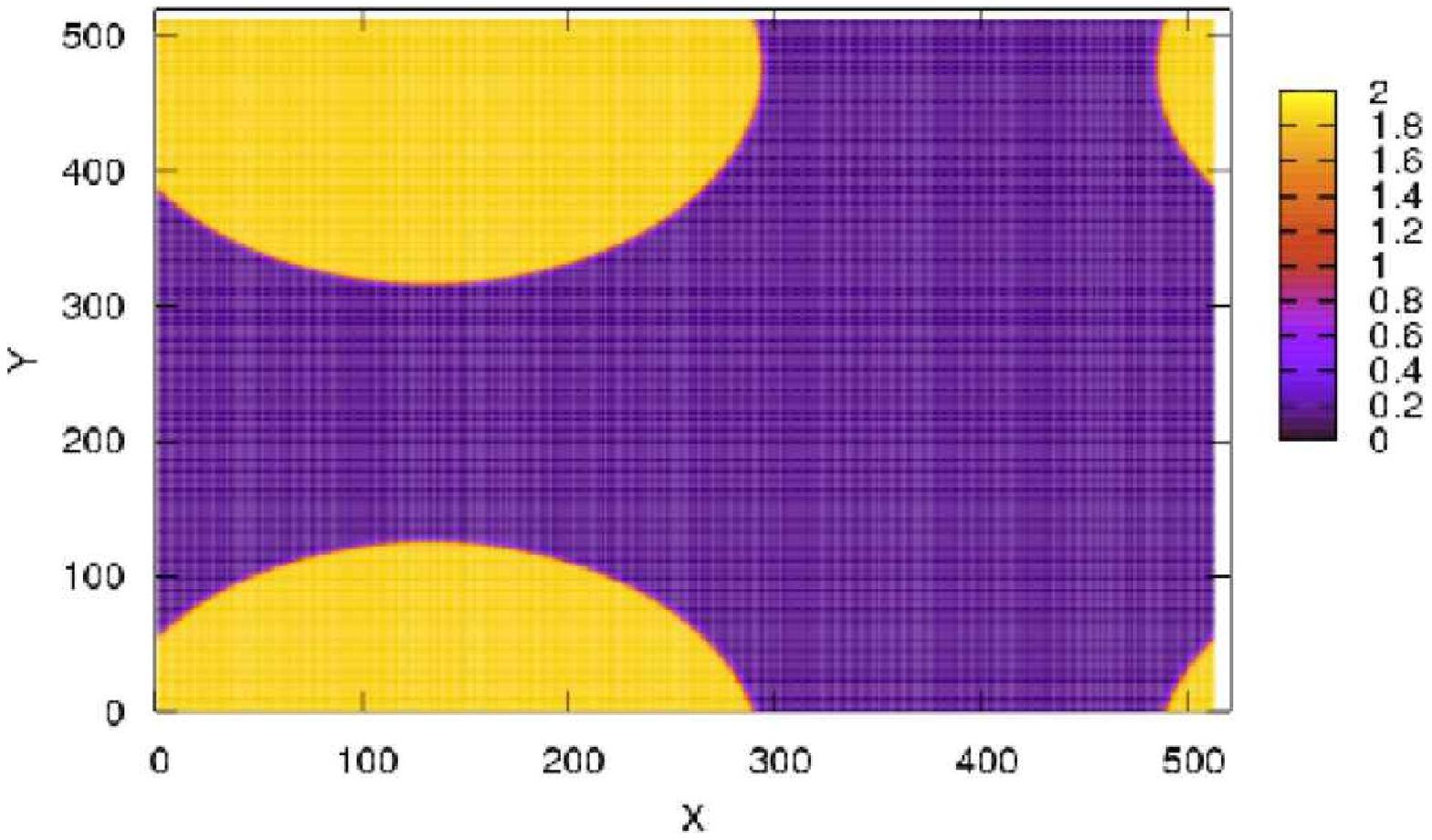}
\includegraphics[scale=0.32]{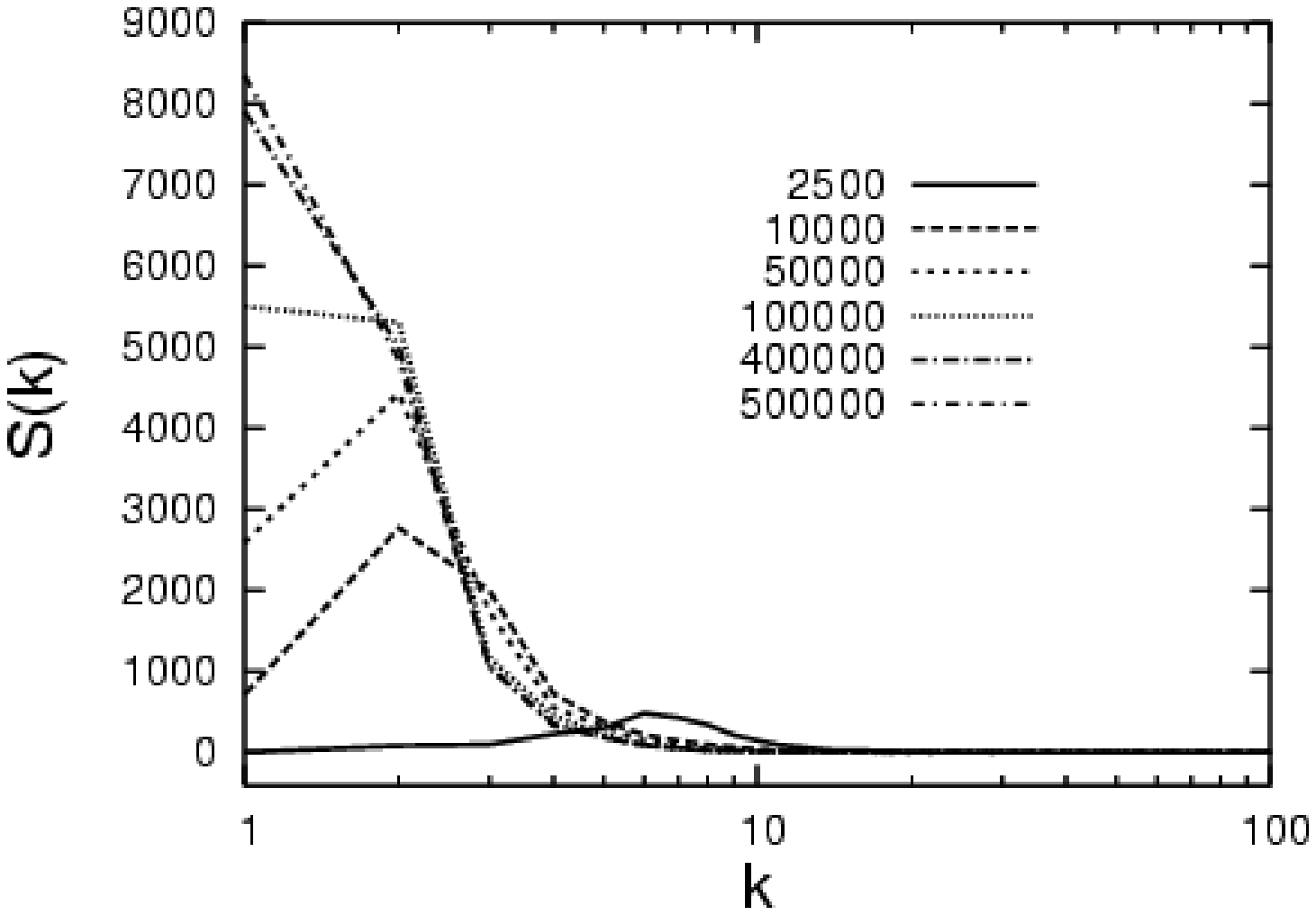}\\
(b)
\includegraphics[scale=0.28]{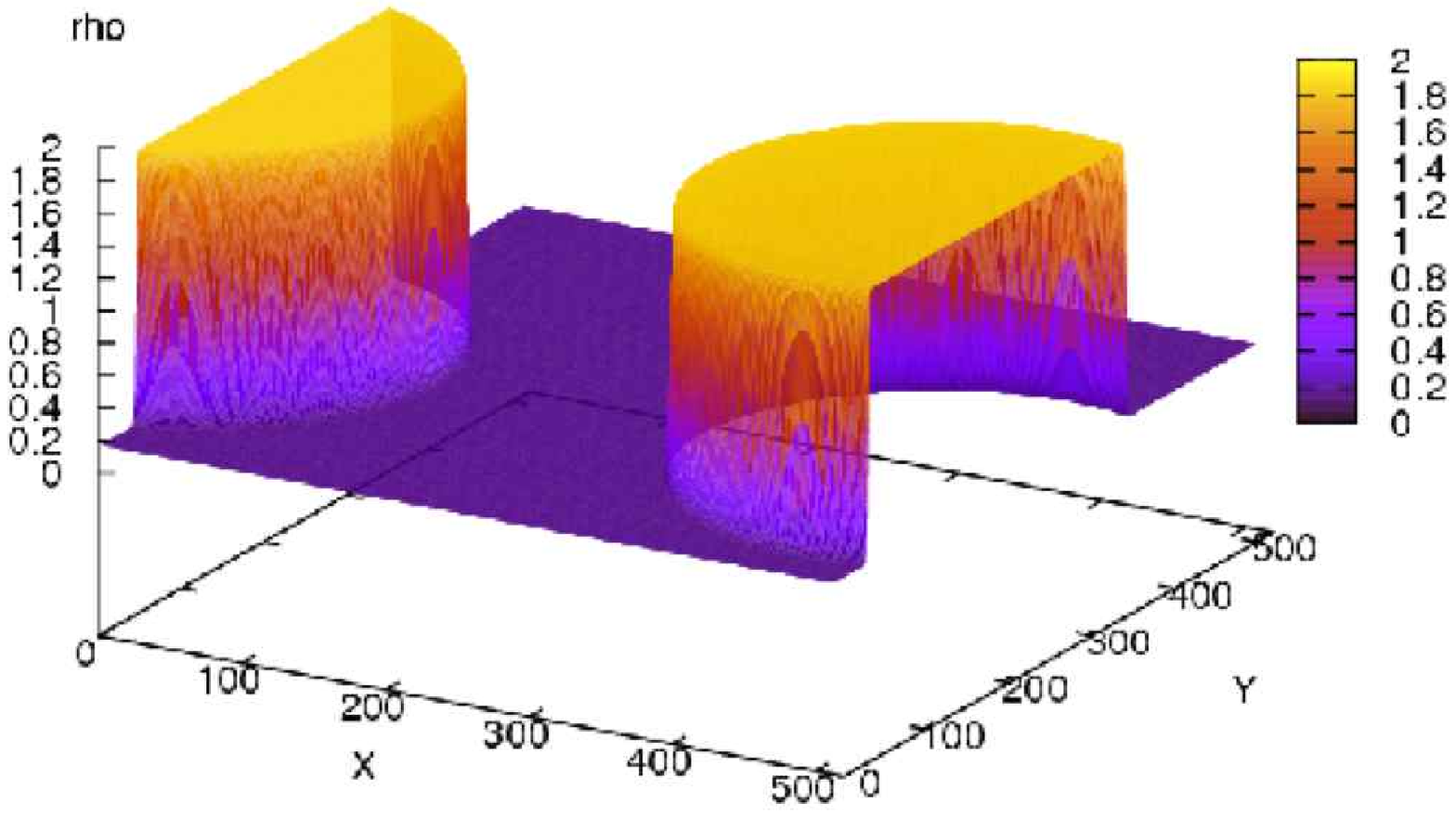}
\includegraphics[scale=0.28]{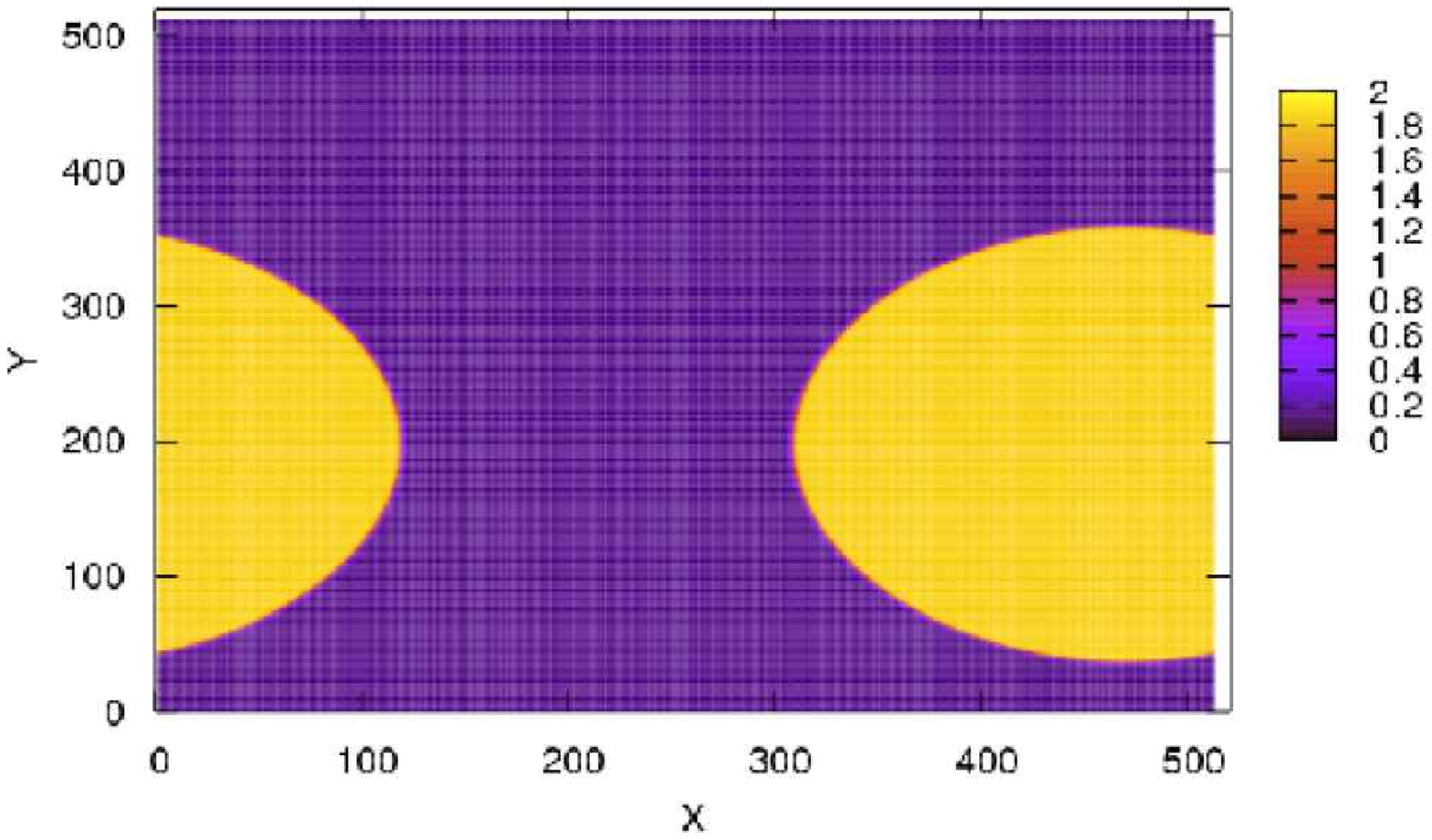}
\includegraphics[scale=0.32]{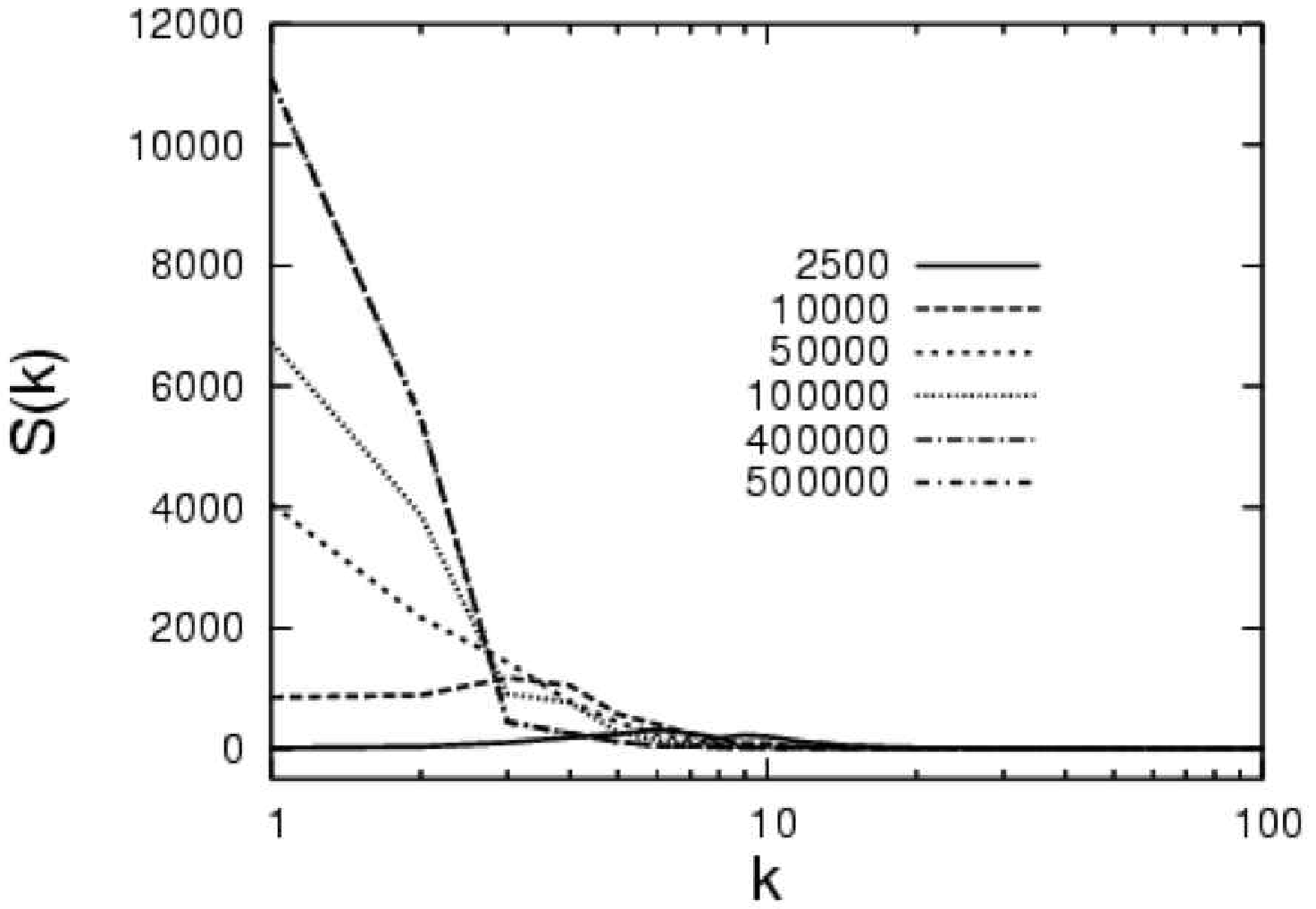}\\
(c)
\includegraphics[scale=0.28]{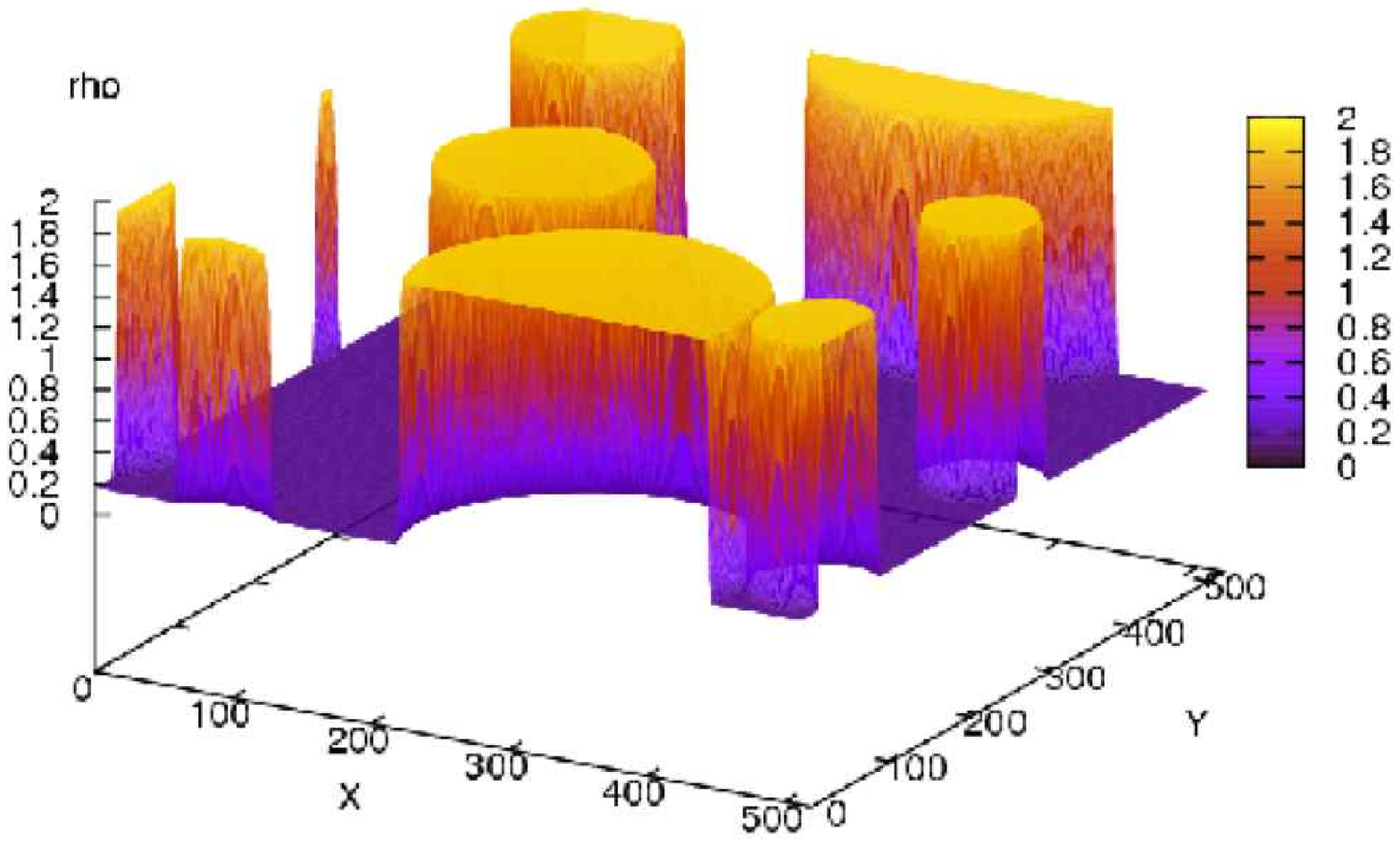}
\includegraphics[scale=0.28]{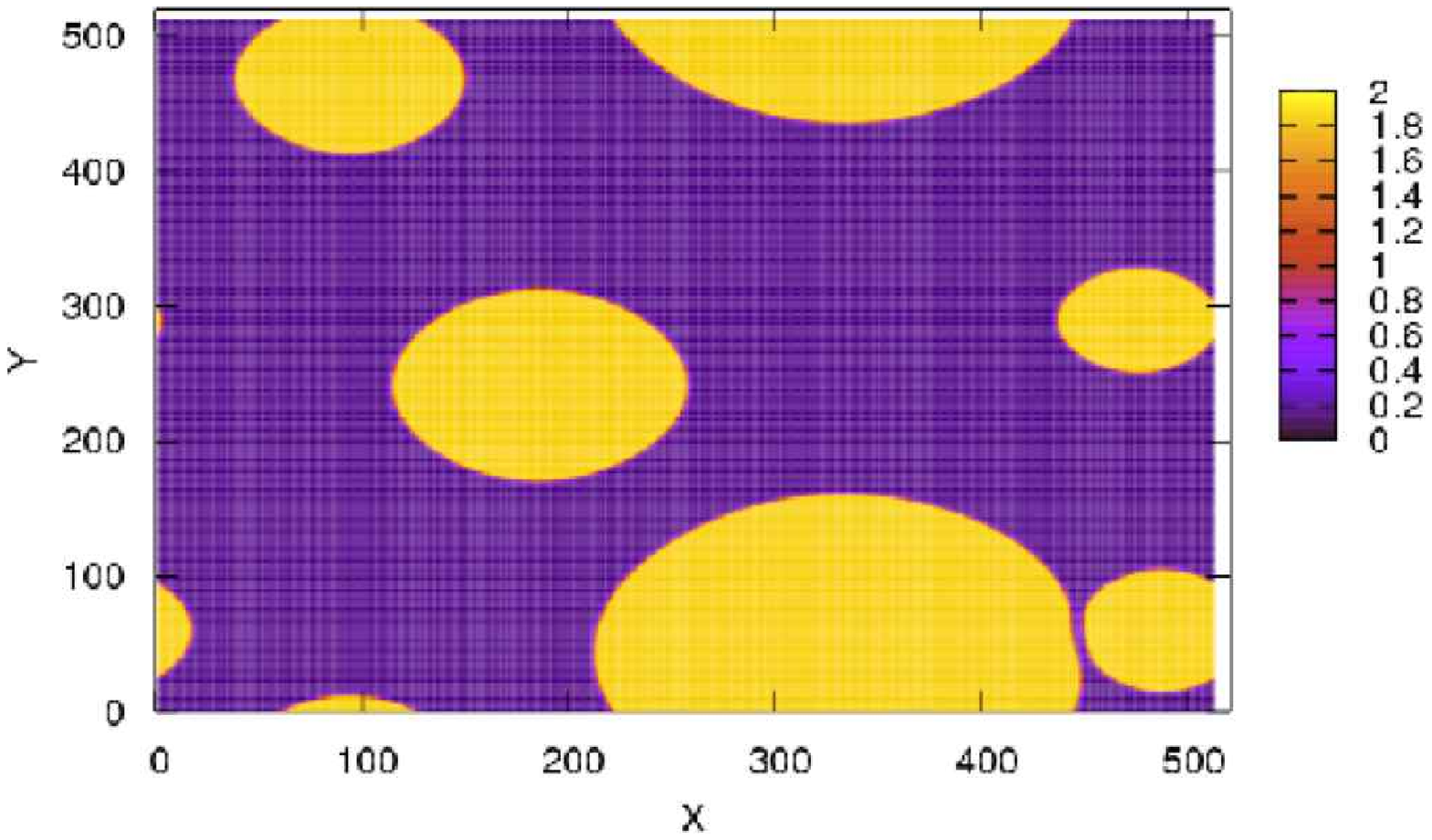}
\includegraphics[scale=0.32]{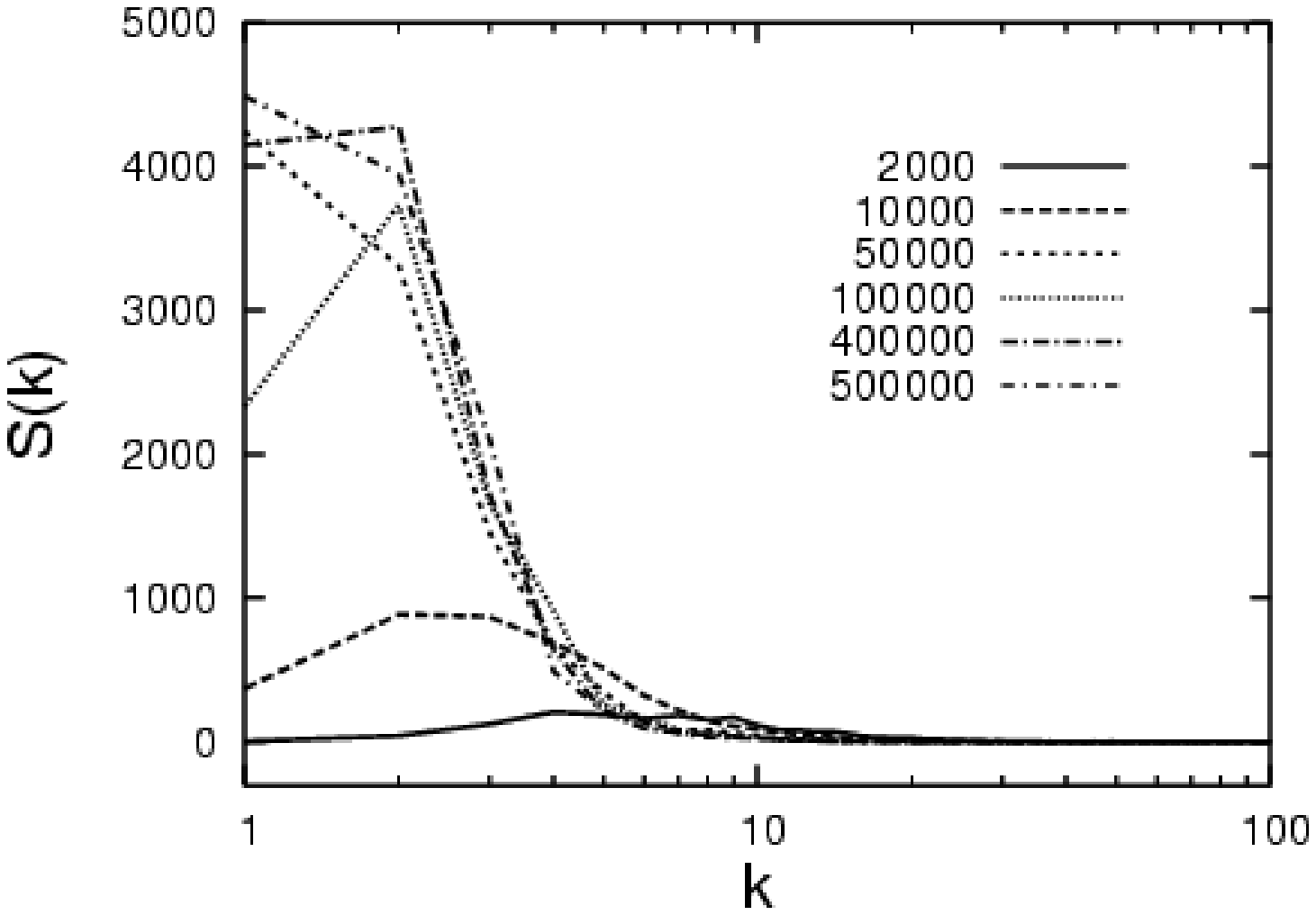}\\
(d)
\includegraphics[scale=0.28]{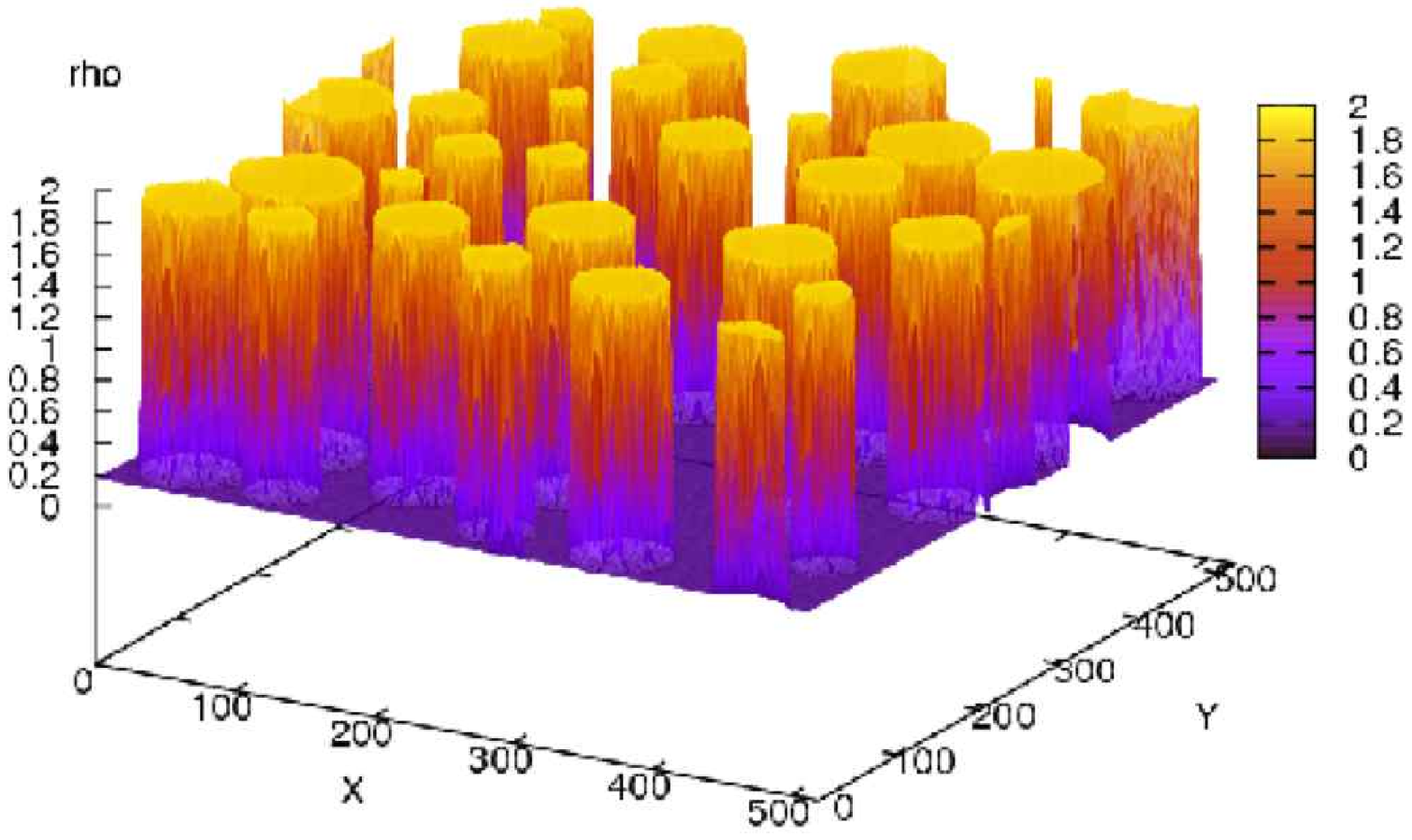}
\includegraphics[scale=0.28]{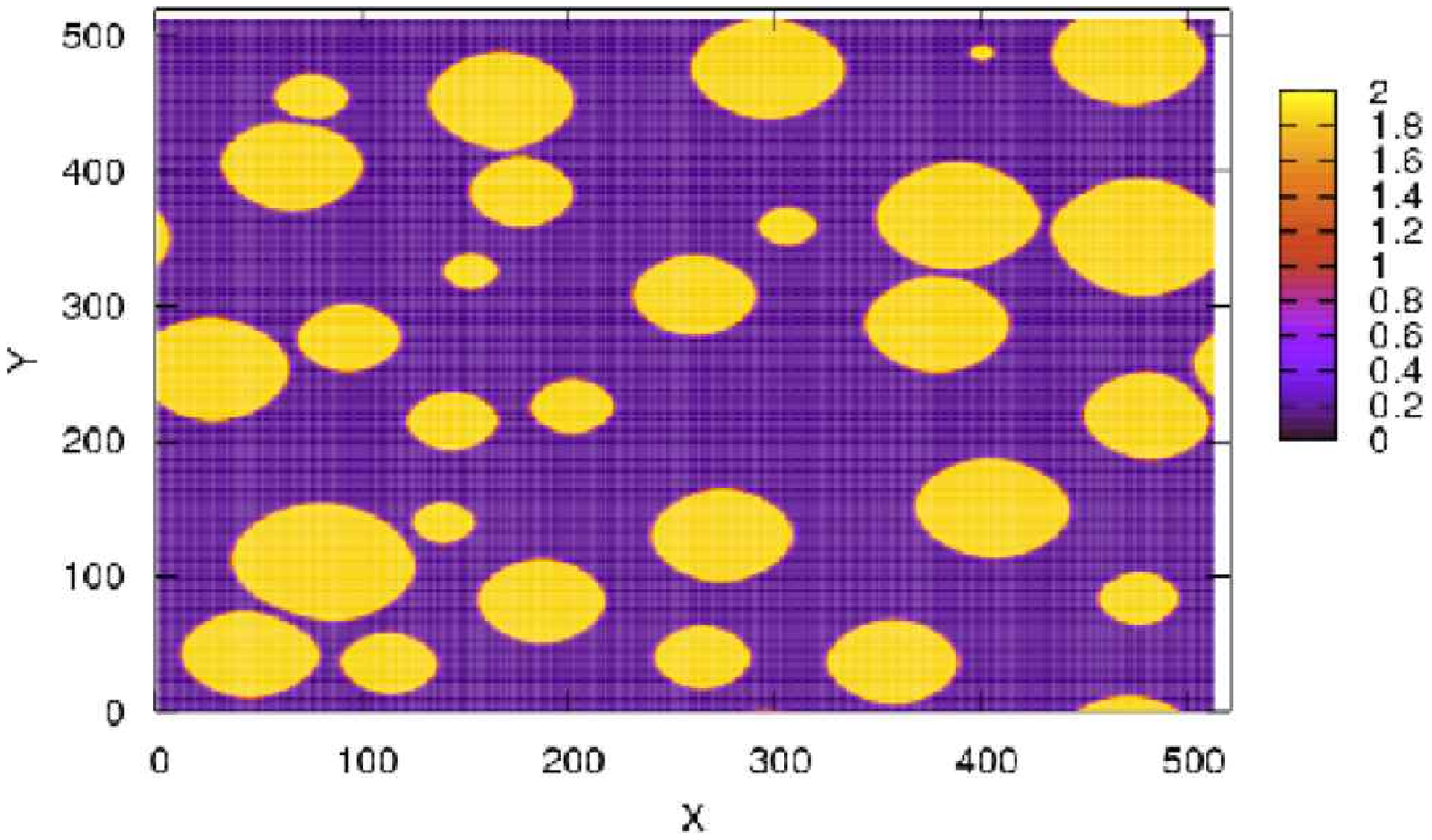}
\includegraphics[scale=0.32]{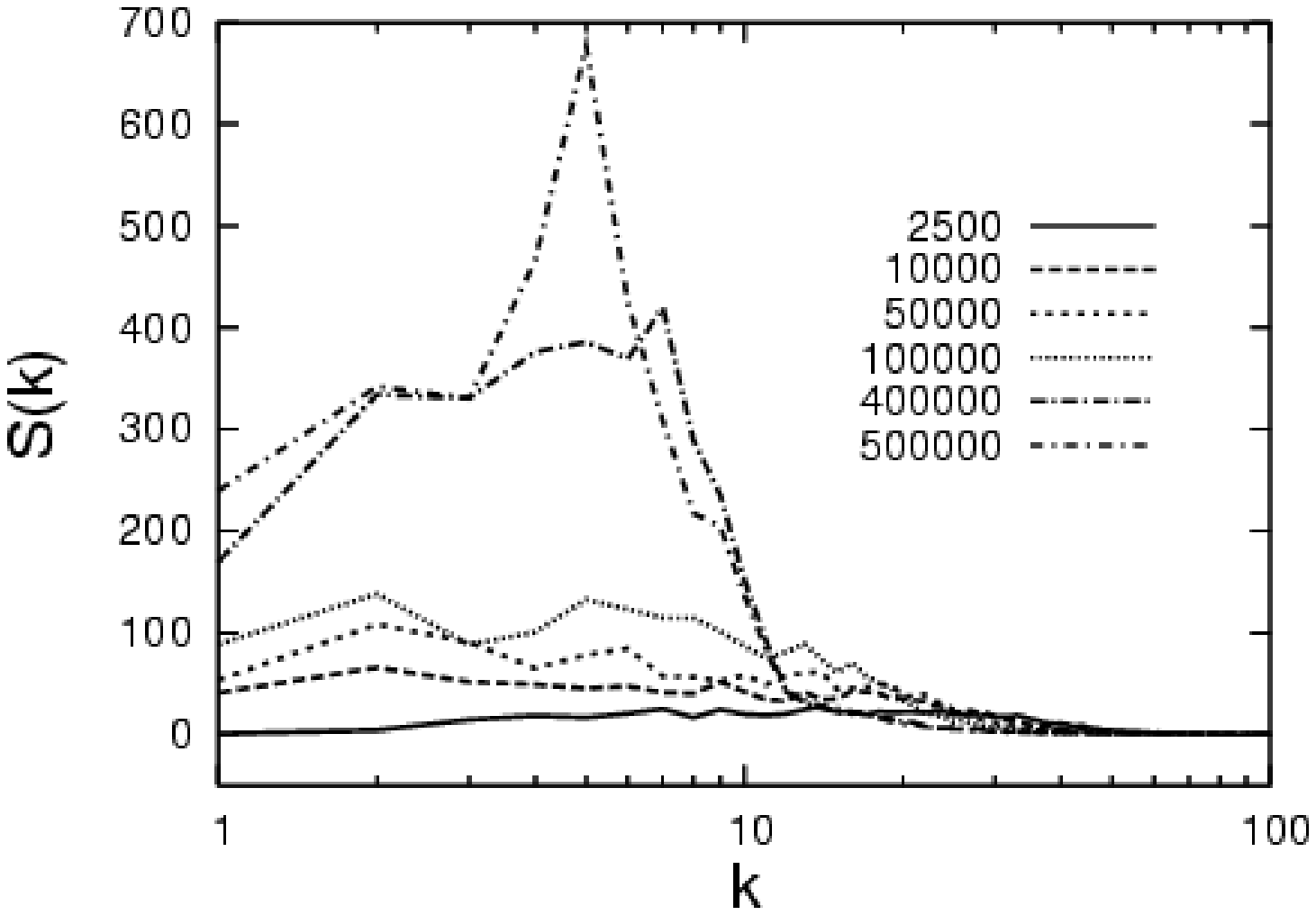}\\

\caption{\small{(Color online) Spatial distribution of the fluid density.
The formation of a large number of droplets with increasing $A_2$ is well visible.
[a]: Standard Shan-Chen, $A_1=-4.9$, $A_2=-4.9$ $nx=ny=512$, $t=500 000$;
[b]: Multi-droplet, $A_2=-2.85$, $nx=ny=512$, $t=500 000$;
[c]: Multi-droplet, $A_2=-0.8$, $nx=ny=512$, $t=500 000$;
[d]: Multi-droplet, $A_2=0.15$, $nx=ny=512$, $t=500 000$.
The right panel shows the Fourier spectrum of density fluctuations. Such spectrum, initially a 
\textit{white noise}, evolves towards a shape peaked at the (inverse) size of the droplets. These Fourier 
spectra show that small-scale contribution is significantly higher when increasing the mid-range repulsion,
that is $A_2$, indicating the formation of long-lived metastable states in the form of small droplets. In particular in the last picture (d), at the end of the simulation there is a clear peak at $R\sim L/2k\approx 30$.}}
\label{Fig:1}
\end{figure}

\begin{figure}

\includegraphics[scale=0.4]{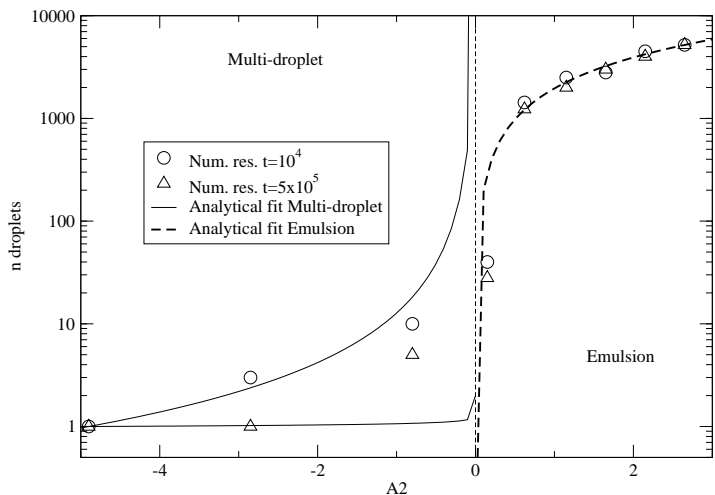}
\caption{\small{Number of droplets versus $A_2$ after a short time $t=10000$ and final time $t=5\times 10^5$. 
The vertical line denotes the transition zone from the Multi-Droplet to the Emulsion 
region, in correspondence with the theoretical spinodal point $A_{2c} = 0$.
The standard SC single-droplet region is associated with $A_2\rightarrow -4.9$ ($G_2\rightarrow 0$).
For $0<G_2<G_{2c}$ metastable multi-droplet configurations are found,
which tend, nevertheless, to the single-droplet equilibrium configuration after a sufficient long time.
For $A_2>0$, the relaxation time associated to the decay to this equilibrium state becomes formally infinite (no changes in time for all observables), indicating that the non-equilibrium phenomena that sustain these metastable states 
experience very slow dynamics.
The solid lines represent two different fits for the two regions, the multi-droplet and the emulsion one. 
Respectively, they are given by: $n(t)=(1-\xi)^{-p(t)}$, with $p(t)=\frac{2}{1+t/10t_{cap}}$ and $\xi=G_2/G_{2c}$;
$n(\xi)=a \xi$ with $a=10000$.
}}
 \label{Fig:2}
\end{figure}

In fig. \ref{Fig:1}, some snapshots of the density at final time are shown for different value of $A_2$.
In fig. \ref{Fig:2}, it is shown the number of droplets, at the end of the simulation, 
as a function of $A_2$.
The simulations show a threshold in phase-separation as $G_2$ increases towards a critical value, $A_{2c}$: 
beyond $A_{2c}$, the density field exhibits numerous stable droplets, distributed according to a quasi-ordered 
configuration, somehow reminiscent of a crystal-like configuration with defects.
The numerical value of $A_{2c}$ can be roughly estimated by noting that, to
fourth-order in the lattice spacing, the total force due to intermolecular 
interactions, $\vec{F_{tot}}=\vec{F}_s+ \vec{F}_m$, is given by:
\be
\vec{F_{tot}}= -\left(  c_s^2 A_1\psi \vec\nabla \psi + \frac{A_2 c_s^4}{2}\ \psi \vec \nabla \Delta \psi
 \right)
\label{eq:ftot}
\ee
where, as previously mentioned, $A_1=G_1+G_2$ controls the magnitude of the phase separation (liquid to gas
density)
and $A_2 = G_1 + \lambda G_2$ is directly linked to the surface tension.
A dimensional argument gives $l^2 = \frac{c_s^2}{2} \frac{A_2}{A_1}=\frac{1}{6} \frac{A_2}{A_1}$,
thus yielding 
\be
l \sim \frac{1}{\sqrt{6}}\sqrt{\frac{A_2}{A_1}} \quad \text{that is,} \quad \frac{l}{l_1}= \sqrt{1-\frac{G_2}{2|G_{eff|}|}}
\label{eq:l}
\ee
where $\lambda=3/2$ has been used in the rightmost expression.
In the above, $l$ is the typical size of a nucleus and $l_1$ is the typical 
single-droplet size for the Shan-Chen case.
With this choice of $\lambda$, the resulting spinodal value, at which $l \rightarrow 0$, turns out to be $G_{2c} = 9.8$, 
corresponding to $A_{2c} = 0.0$.
For this value, the coefficient in front of the second term in Eq. (\ref{eq:ftot}) vanishes, thus signaling 
the onset of a phase-transition.
This value is found to be in good agreement with the numerical simulations, which indicate 
complete nucleation starting around a value of $A_2 \approx 0$, as shown in Fig. \ref{Fig:2}.
The number of droplets in the first
region, called Multi-droplet region, can be described as a function of time 
as $n(t)=(\xi-1)^{-p(t)}$, with $p(t)=\frac{2}{1+t/10t_{cap}}$, where $\xi=G_2/G_{2c}$
and $t_{cap}=\frac{H\mu}{\gamma}$. 
This relation can be related to simple statistical physics 
arguments \cite{Chi_07}.
The region after the transition, where nucleation takes place, has been 
fitted by a simple linear
function $n(\xi)=a\xi$, where $a=10^4$.
It is worth mentioning that the same linear behaviour in the 
emulsion region is also observed for a coarser domain \cite{Chi_07}.
However, the coefficient $a$ is not universal, as it depends on the domain size.
This may be related to the breakdown of scale invariance of phase-separating fluids as observed in \cite{Wag_98}.

\begin{figure}
(a)
\includegraphics[scale=0.4]{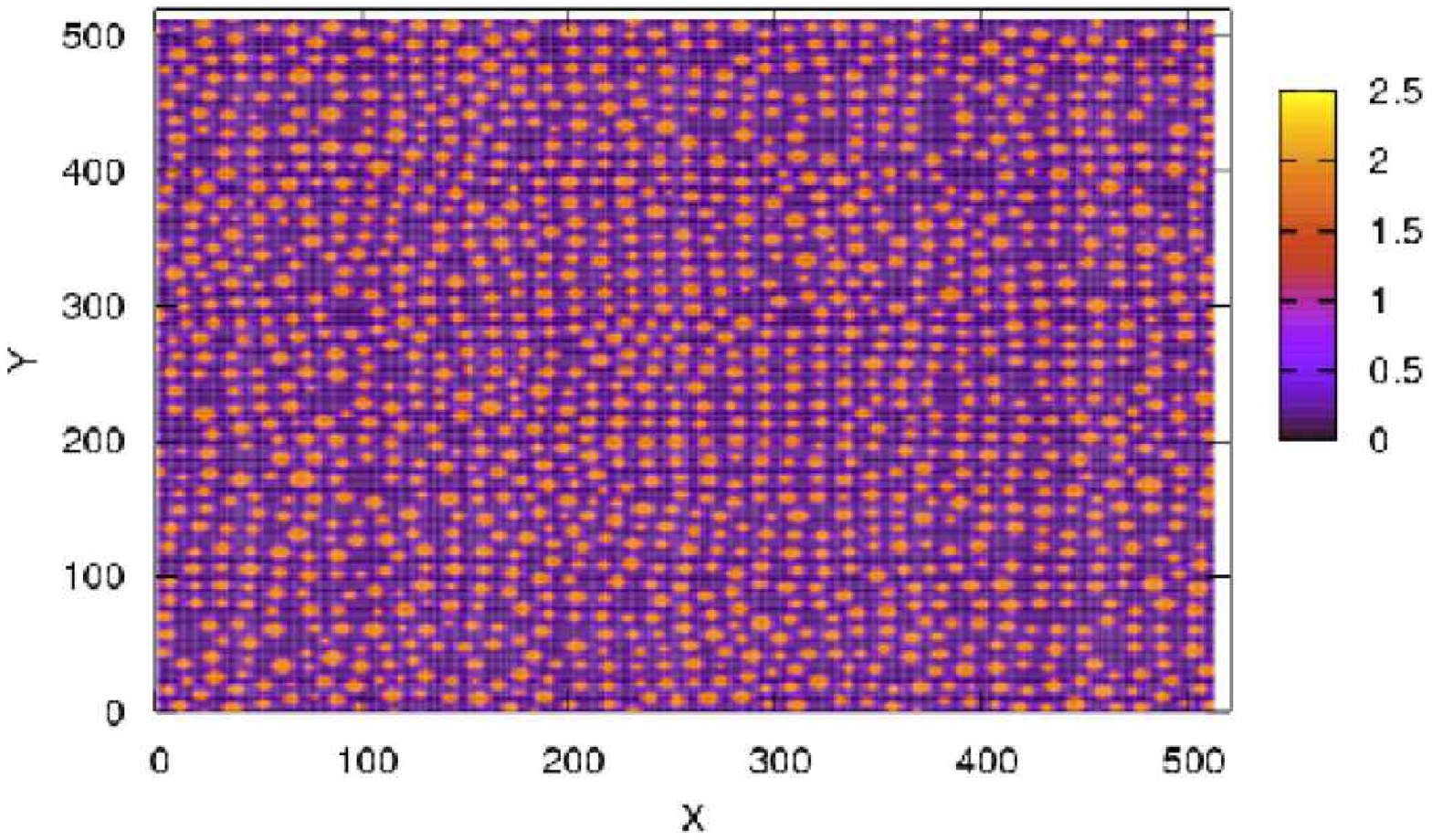}
\includegraphics[scale=0.4]{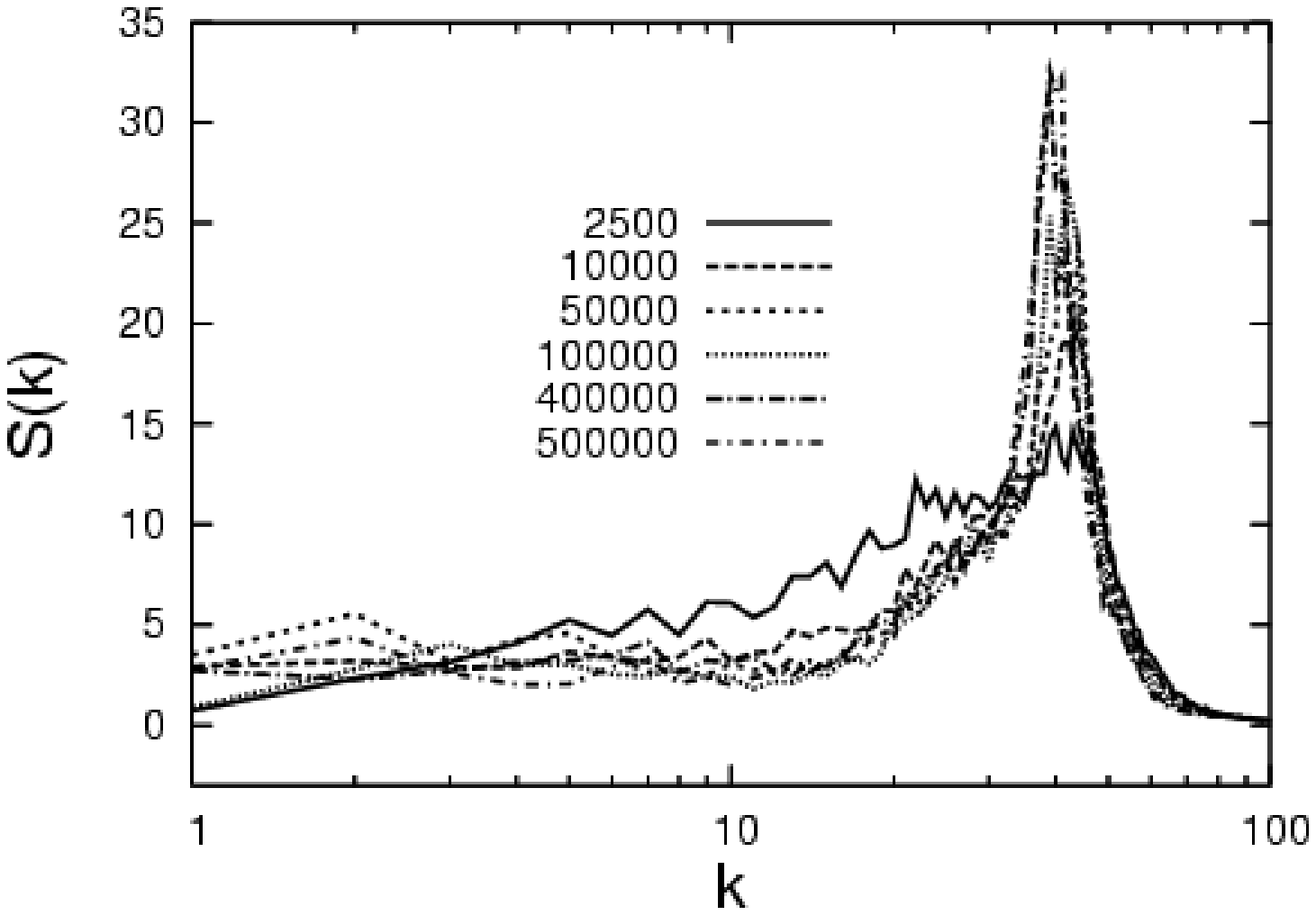}\\

(b)
\includegraphics[scale=0.4]{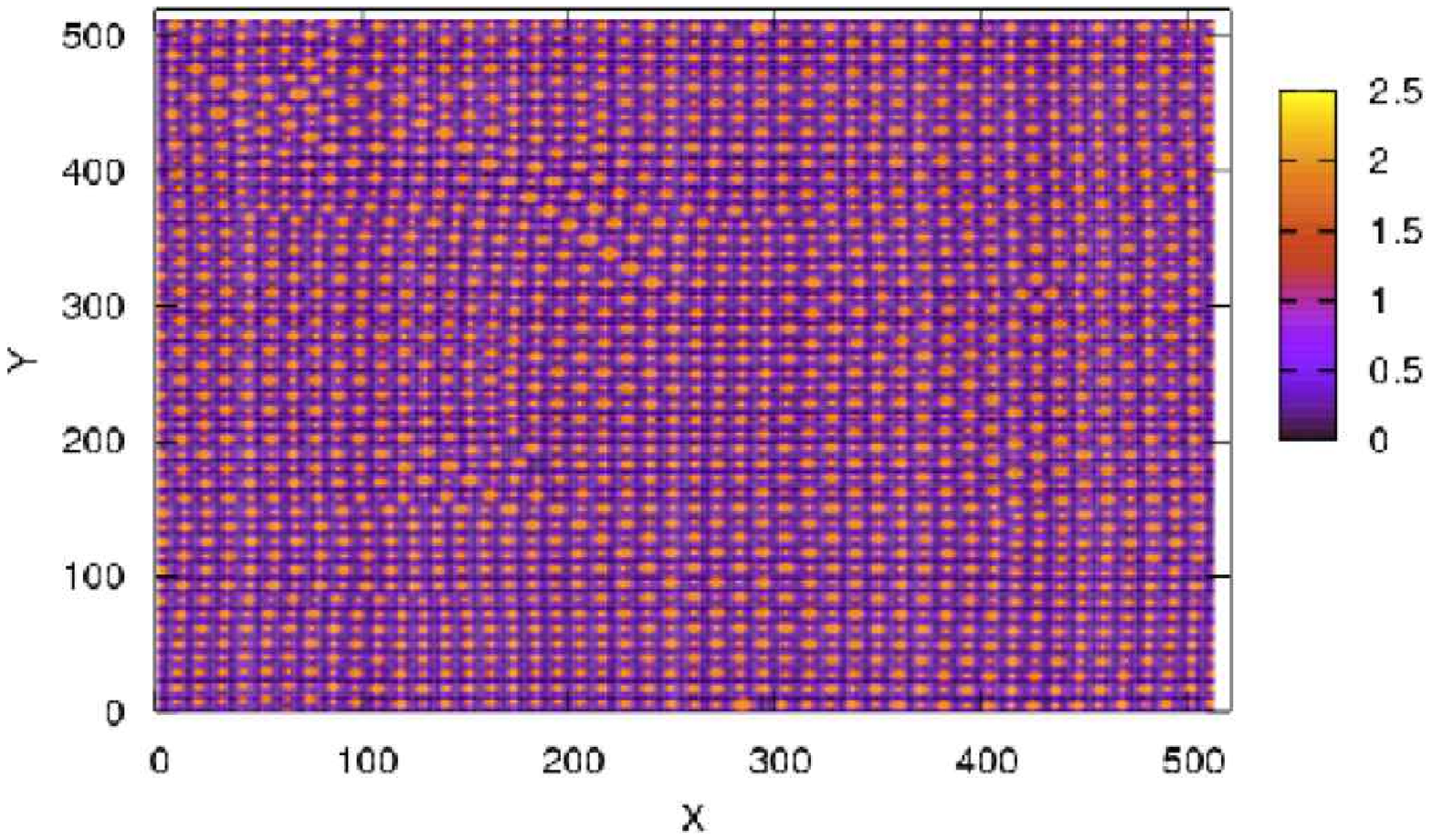}
\includegraphics[scale=0.4]{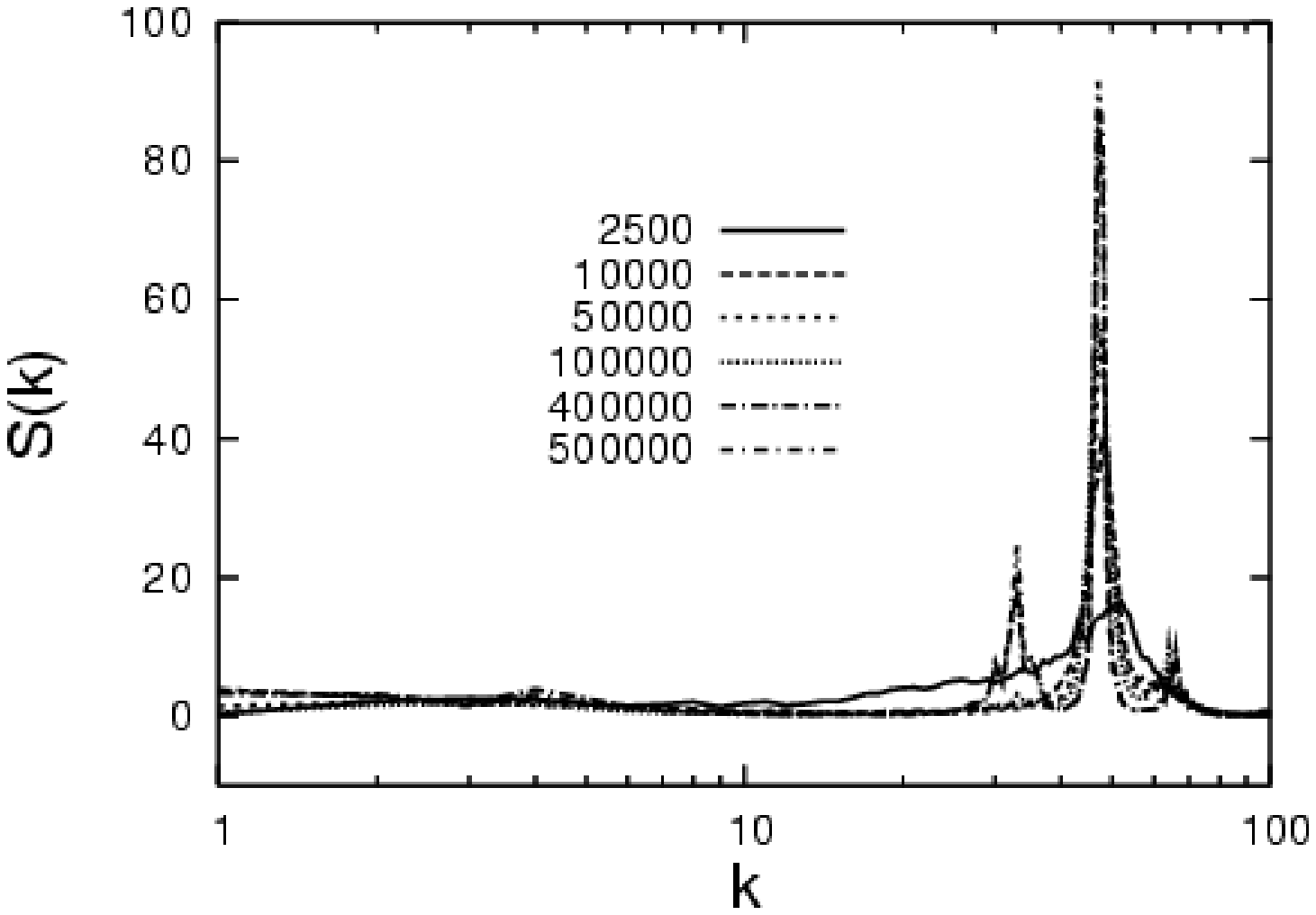}\\

\caption{\small{(Color Online) Spatial distribution of the fluid density for the spray-emulsion configuration.
The crystal-like ordered structure of droplets is evident in fig. (b), where most of droplets are organized
into 6-neighbourhood structures.
 $A_1=-4.9$ in all cases.
[a]: $A_2=0.65$, $nx=ny=512$, $t=500 000$;
[b]: $A_2=1.15$, $nx=ny=512$, $t=500 000$. The spectrum of density fluctuations shows a sharp
peak corresponding to the typical size of the droplet.
For the case (a) we obtain a typical radius of $R\approx 8$, whereas in the second and more ordered configuration, $R\approx 6.5$.}}
\label{Fig:3}
\end{figure}
\begin{figure}
(a)
\includegraphics[scale=0.4]{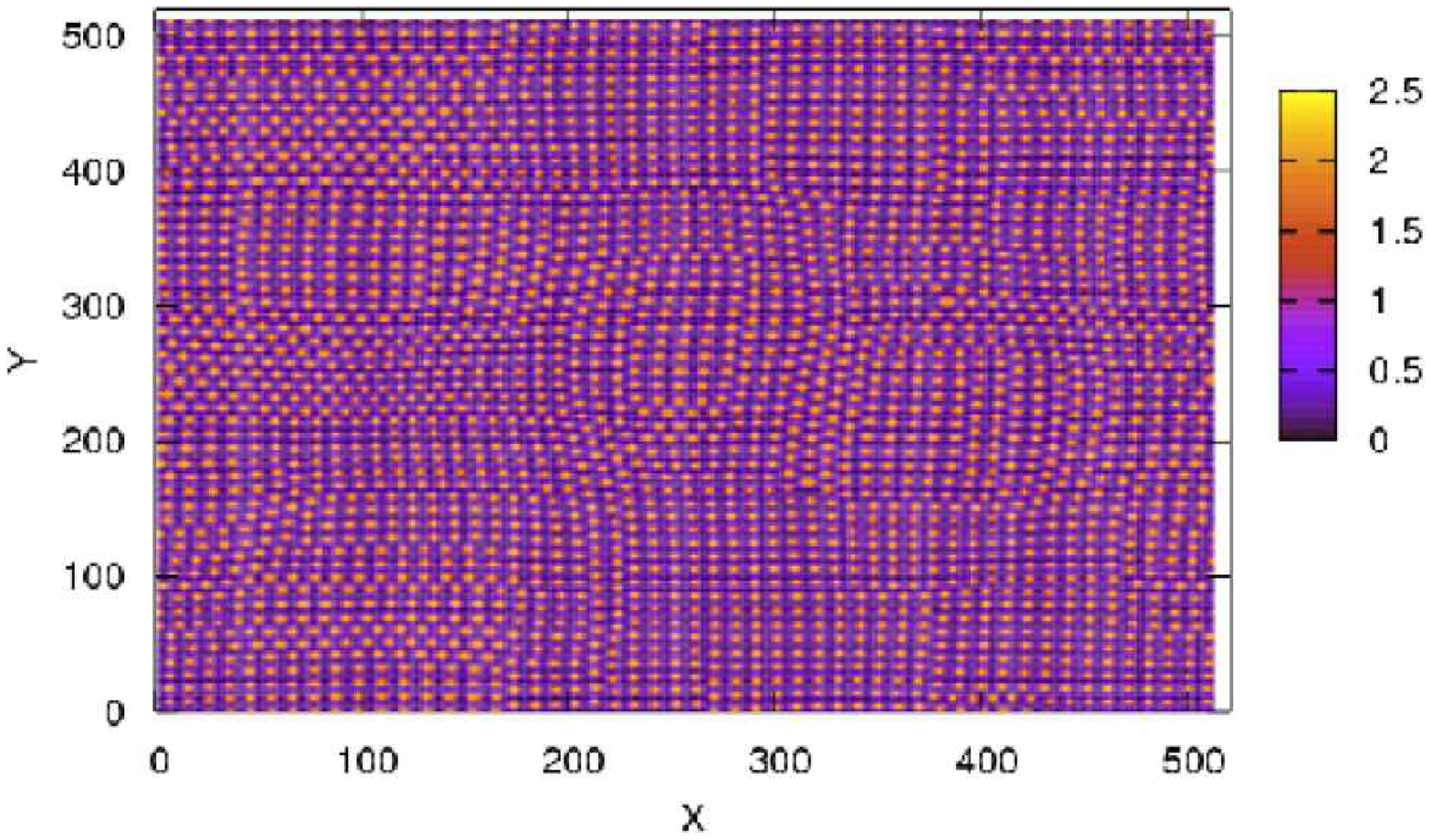}
\includegraphics[scale=0.4]{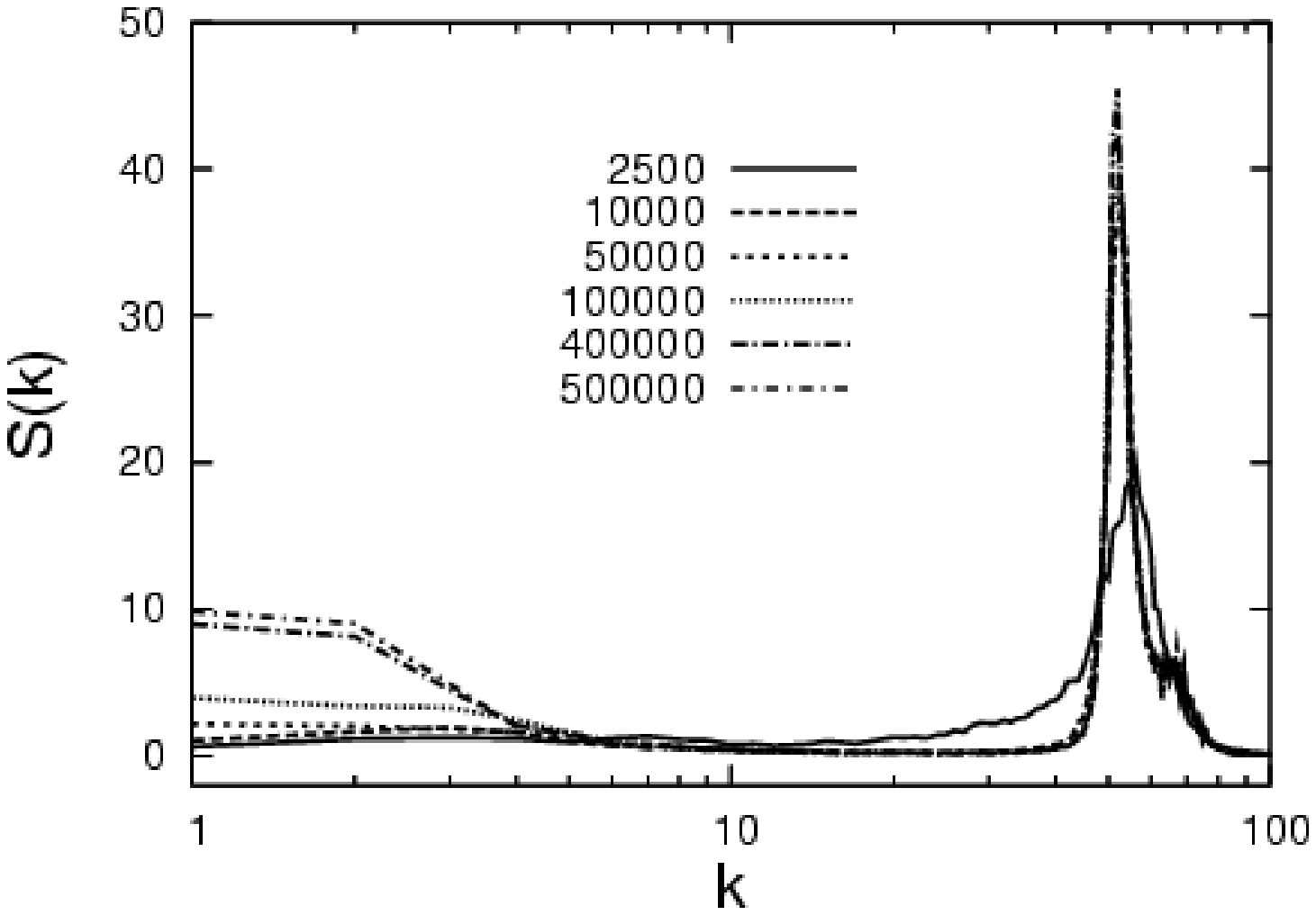}\\
(b)
\includegraphics[scale=0.4]{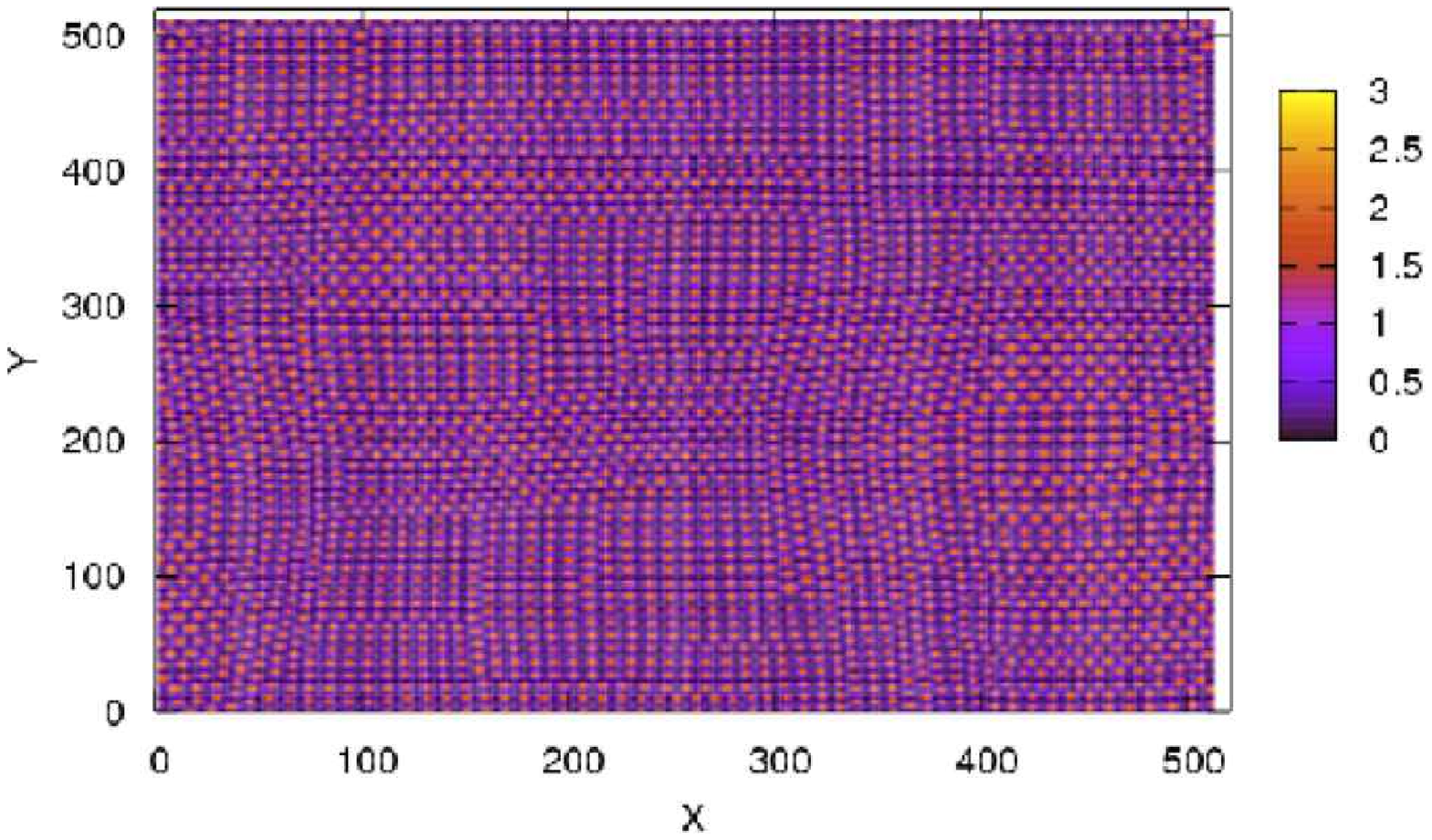}
\includegraphics[scale=0.4]{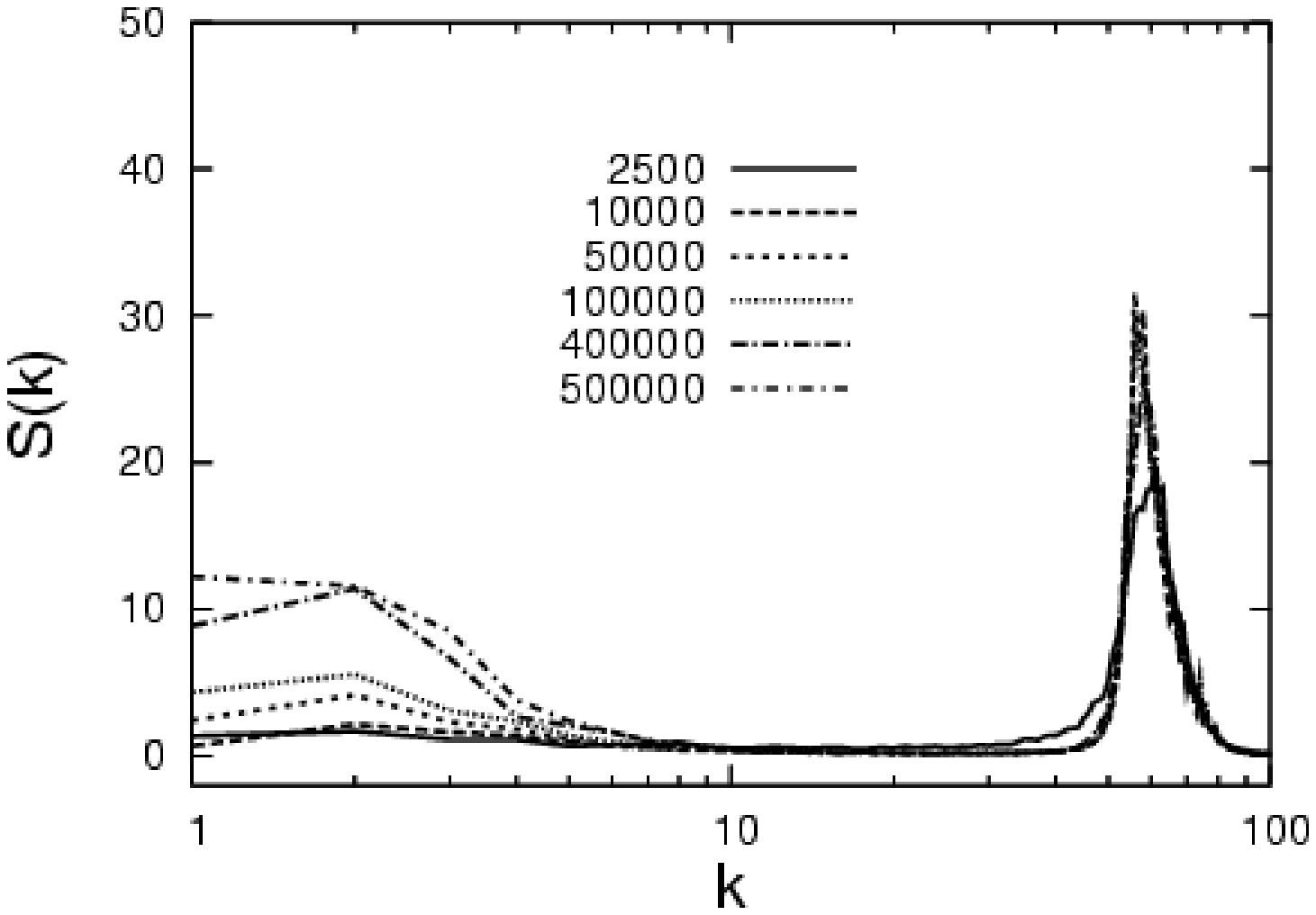}\\
(c)
\includegraphics[scale=0.4]{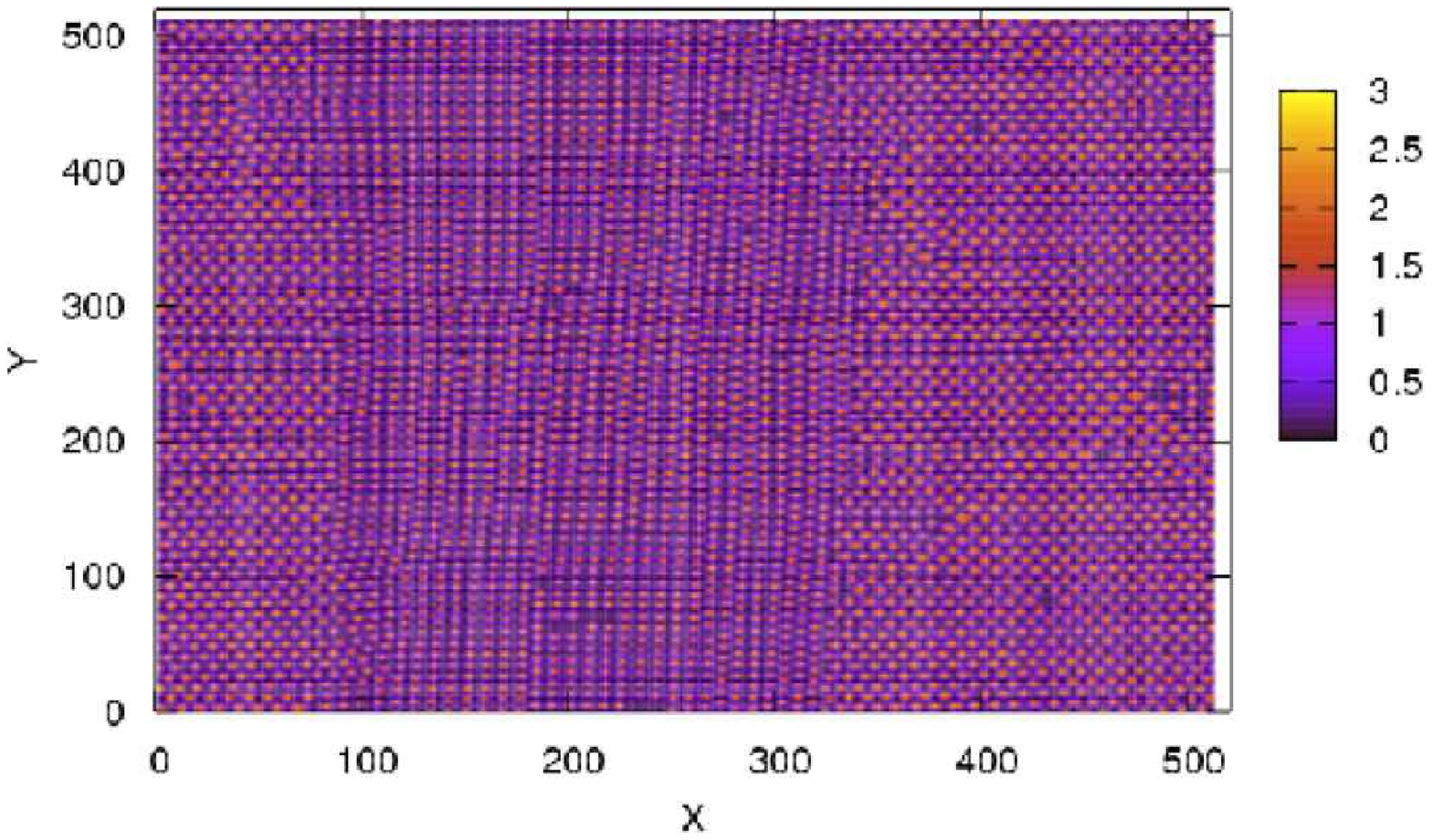}
\includegraphics[scale=0.4]{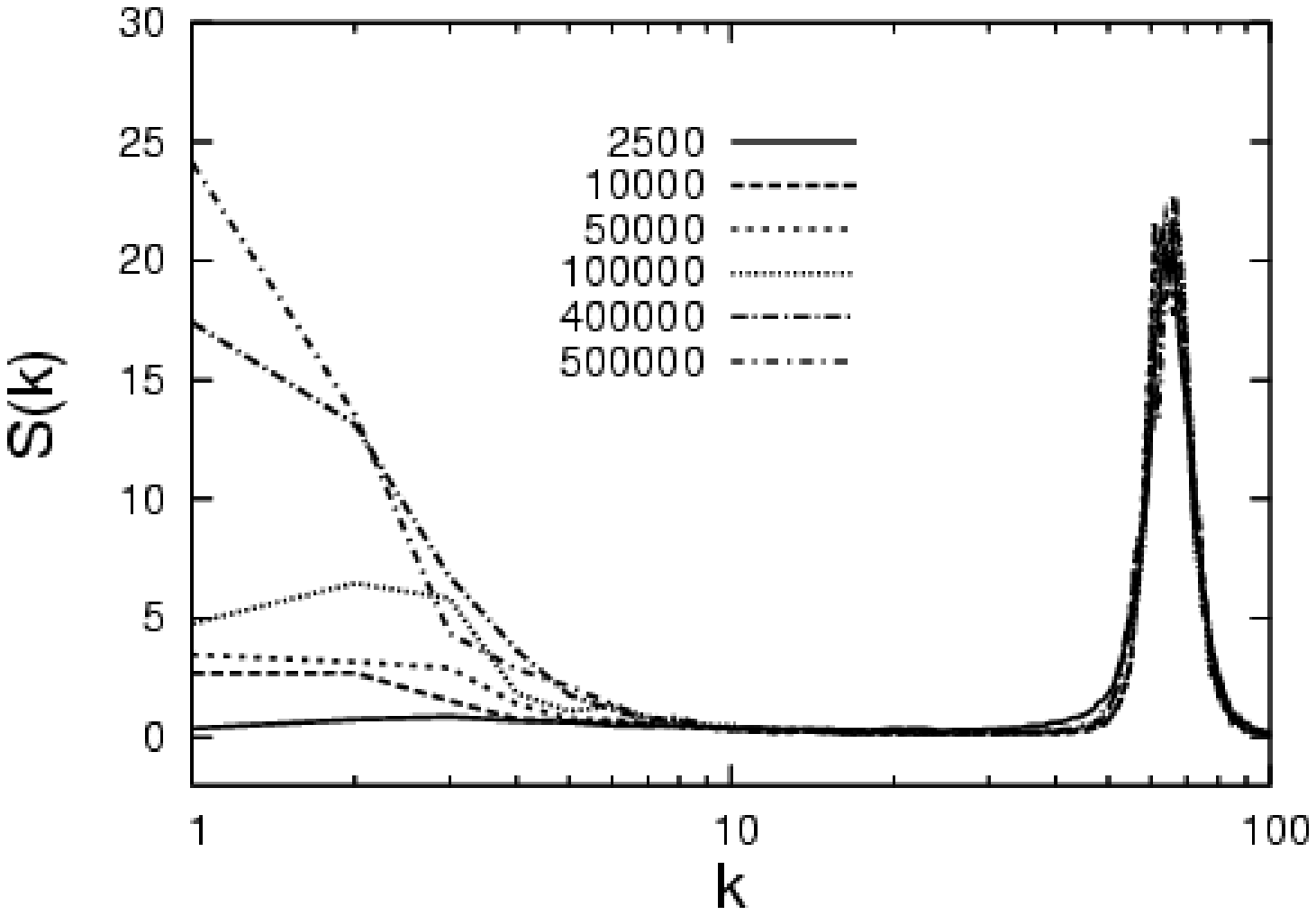}\\

\caption{\small{(Color online) Spatial distribution of the fluid density for the spray-emulsion configuration.
As in previous picture fig. \ref{Fig:3}, $A_1=-4.9$ in all cases and 
the corresponding Fourier spectra of density fluctuations are reported in the right panel. 
[a]: $A_2=1.65$, $nx=ny=512$, $t=500 000$;
[b]: $A_2=2.15$, $nx=ny=512$, $t=500 000$;
[c]: $A_2=2.65$, $nx=ny=512$, $t=500 000$. 
Besides the sharp peak centered around the mean size of the droplets, the build-up of a low-$k$ component with increasing $A_2$ is well
visible, corresponding to the formation of large-scale domains
indicating a higher degree of order in the global structure. 
The configuration presented in (c) is strongly reminiscent
of a crystal, with very few defects. 
For these cases, the typical radius is estimated as follows:
(a) $R\approx5.1$; (b) $R\approx4.4$; (b) $R\approx3.9$.}}

 \label{Fig:4}
\end{figure}


It is instructive to inspect the spatial distribution of the phase-separated fluid
as the $A_2$ parameter is increased.
In the present model, as well as in the standard Shan-Chen, phase separation starts immediately and spontaneously, once the 
parameters are chosen in the critical range: in the Shan Chen model 
and in the two-belt (with $A_2$ below the \textit{critical value}), small droplets coalesce in larger droplets
of increasing size, until only a few of them, or even just one, are left. 
This is the spatial configuration which minimizes the surface energy expenditure.
To study this spontaneous coalescence and its relation to the model parameters, a Fourier analysis 
of the density field has been conducted, based on the structure factor S(\textbf{k},t):
\be
  S(\textbf{k},t)=\frac{1}{N}\bigg\vert \sum_{\textbf{x}} \ \big[ \rho(\textbf{x},t) - \overline{\rho}(t) \big] e^{i\textbf{k}\cdot\textbf{x}} \ \bigg\vert ^2
\ee
where $\textbf{k}=(\frac{2\pi}{L})(l,m)$, $\textbf{x}$ is the lattice point, $L$ is the linear lattice size ($=512$ in our case), $N = L^2$ is the total number of grid   points; $\rho(\textbf{x},t)$ is the density field at time $t$ and $\overline{\rho}(t)$ is the average density field at time $t$.
It is possible to average the structure factor in \textbf{k} space, as follows:
$S(k,t)=\frac{\sum_{k} S(\textbf{k},t)}{\sum_{k} 1}$,
where the sum is over a circular shell defined by $(n-1/2) \le |{\bf k}|L/2\pi < (n+1/2)$.

This first moment of the circularly-averaged structure factor can then be used to 
assess the characteristic length scale of the droplet, $R(t) = 2\pi/\overline{k}(t)$, where 
\be 
  \overline{k}(t) = \frac{\sum_{k} k S(k,t)}{\sum_{k} S(k,t)}
\label{eq:rad}
\ee
Note that $k=1$ means $R=\frac{L}{2}$.
The right columns in Figs. \ref{Fig:1}, \ref{Fig:3} and \ref{Fig:4} show the time evolution of length scales for 
various $(A_1,A_2)$. 
As is well visible from the figures, after $500000$ time steps, all configurations have 
settled down to their steady state, except the configuration with $A_2\approx A_{2c}$ the typical droplet size
being a decreasing function of $A_2$. 
It is interesting to notice the growth of macroscopic islands, cutting
across the entire computational domain, in the emulsion region.
This is reflected by a significant build-up of the low-$k$ region 
of the spectrum, yet another signature of a phase-transition behaviour.
\begin{figure}

\includegraphics[scale=0.4]{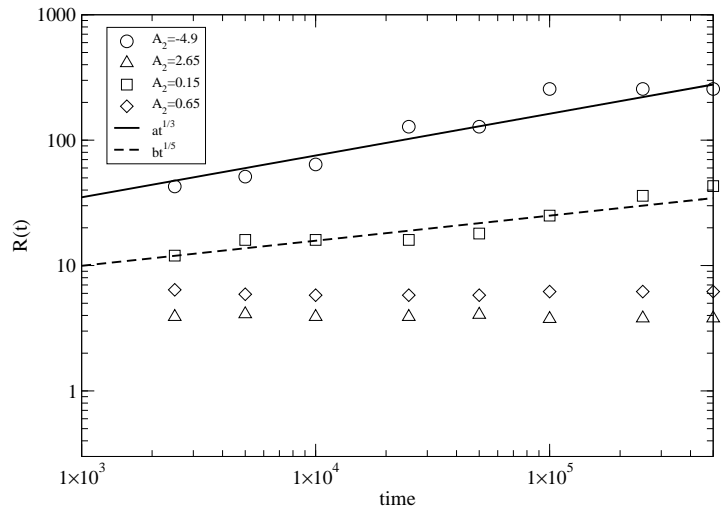}
\caption{\small{Time evolution of average domain size $R(t)$ (lattice units) versus time (time steps) for different cases. Curves from top to bottom correspond to systems with increased mid-range repulsion ( systems with increased average ``surfactant'' concentration). The straight line represents a power law $R(t)\sim (t-t_0)^{1/3}$ which is typical of diffusive growth.
The dashed curve is $R(t)\sim (t-t_0)^{1/5}$}}
\label{Fig:radius}
\end{figure}
In figure \ref{Fig:radius}, we show the evolution in time of the typical radius of some configurations.
The radius is calculated from the circularly averaged structure factor, as described by eq. (\ref{eq:rad}).
For the standard Shan-Chen case, the radius grows till the maximal value of $R\sim L/4$, corresponding to a single droplet.
In this case, the domains grow according to a sub-diffusive 
power-law $R(t)\sim(t-t_0)^{\alpha}$, with a growth exponent 
$\alpha=1/3$ \cite{Lif_61}.
In the other cases, after the transition to the emulsion region, the asymptotic radius attains a much smaller value.
For $A_2=0.15$, the radius still grows, although more slowly,
with a growth exponent $\alpha=1/6$, indicating 
that this metastable state will reach the asymptotic 
single-droplet state in a very long, but finite, time. 
On the contrary, for the other two cases in the emulsion region, the radius 
does not show any appreciable change over the entire simulation
time-span. These states appear completely frozen and do not
show any visible dynamics towards a more stable state.

\section{Pseudo-potential energy evolution}

The pseudo-potential LB models bears a formal resemblance to dynamic mean-field Ising formulations 
of magnetic systems. 
Of course, a major difference with respect to Ising systems is that our
fluid model is clearly not a Hamiltonian one.
It is nonetheless of interest to define a pseudo-potential 
energy, $E(t)=E_s(t)+E_m(t)$, where
\be 
  E_s(t) = \frac{1}{2}\sum_{x,y}\psi(x,y;t)\sum_{i=0}^{b_1}\ (G_1 \ w_i + G_2 \ p_{si})\ \psi(\vec{x}_{si};t)
\ee

\be
 E_m(t) = \frac{1}{2} \sum_{x,y} \psi(x,y;t) \sum_{i=0}^{b_2} G_2\ p_{mi}\ \psi(\vec{x}_{mi};t)
\label{eq:e2}
\ee
are the contributions from the first and second belts, respectively.
This definition, suggested by a direct analogy with the 
Ising Hamiltonian $H[s]=\sum_x \sum_{y=x \pm 1} s(y) J(x,y) s(x)$, is
also in line with the expression of the forces, (\ref{PFORCE}).

By expanding $\psi_i$ in powers of $c_i$, to zeroth order (local-density approximation), we obtain
the \textit{bulk} contribution:

\be
\label{Ebulk}
 E_{bulk} = \frac{A_1}{2} \sum_{x,y} \psi^2(x,y)
\ee

while the next order (weak-gradient approximation) delivers a surface term: 

\be
\label{Esurf}
 E_{surf} = \frac{A_2}{2} \sum_{x,y} \left(\nabla \psi(x,y)\right)^2
\ee
where we used the normalizations in Eqs. (\ref{weights_1}) and (\ref{weights_2}).

\begin{figure}
\includegraphics[scale=0.8]{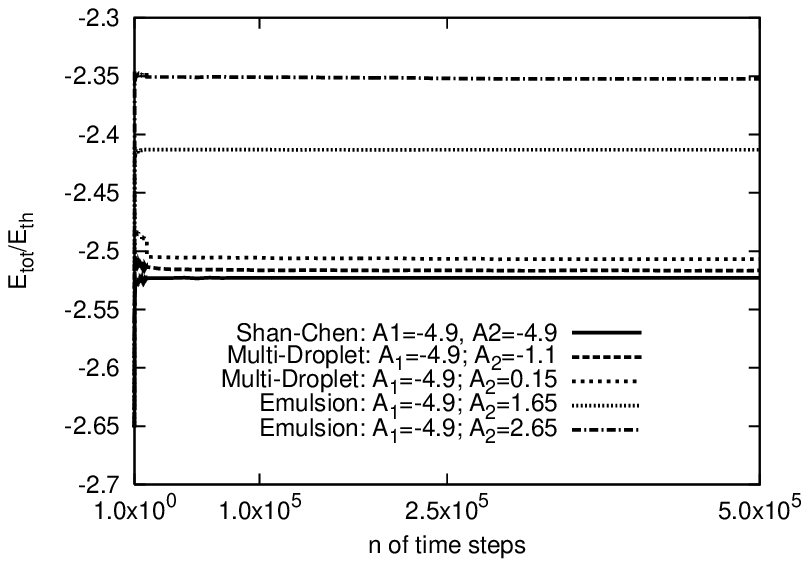}
 \includegraphics[scale=0.8]{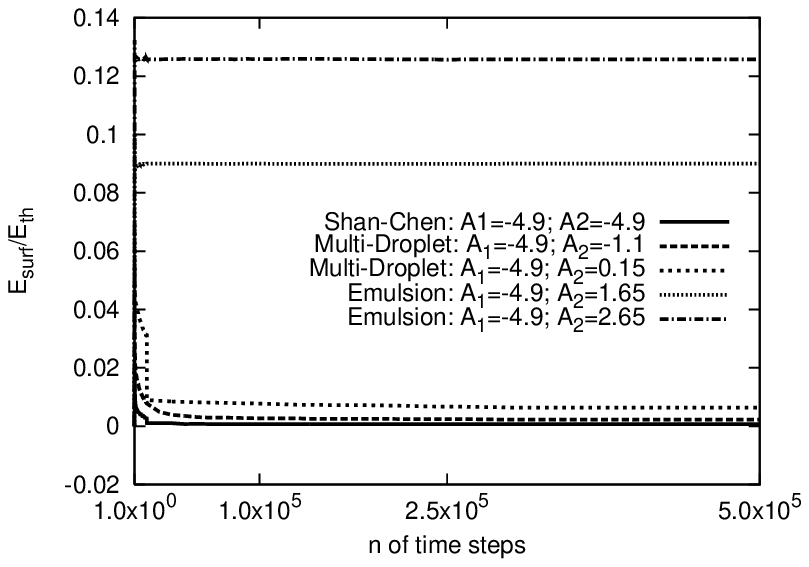}
\caption{\small{(a): Time evolution of total pseudo-energy $E(t)$ in units of the thermal
energy $E_{th} = \rho c_s^2 L^2$ for increasing value of the second belt coupling $A_2$.
The steady-state value of the pseudo-energy is a monotonically increasing function of $A_2$, the
standard SC case ($A_2=0$) being the lowest-energy phase-separated configuration. In that sense,
the standard SC configuration may represent a ground state with a discrete spectrum of excited states triggered by increasing discretely $A_2$. 
(b): Time evolution of $E_{tot}$ and $E_{bulk}$, Eqs. (20) and (22).
The surface energy, given by the difference $E_{tot}-E_{bulk}$ is 
found to be always positive, as it should be. 
Furthermore, it is possible to see that the surface contribution 
increases from the Multi-Droplet to the Emulsion region (see Fig. 3), 
consistently with the picture of states 
which become increasingly excited with increasing $A_2$. 
In the above scales, the two plots in the SC case would almost coincide, since the surface contribution is a mere $\approx 0.001$, instead of $\approx 6\%$ 
for the emulsion case.}}
\label{Fig:5}
\end{figure}

In figure (\ref{Fig:5}), the ratio of the global pseudo-energy to the
thermal energy $E_{th}=\rho c_s^2 L^2$ is shown as a function of time 
for increasing values of the second-belt coupling $A_2$.
The figure shows that the steady-state value of the pseudo-energy is a monotonically 
increasing function of $A_2$, the standard SC case ($A_2=0$) being the lowest-energy phase-separated 
configuration. The initial rise of the global energy reflects the build-up of surface
energy due to interface formation. Once such short transient is settled down, the pseudo-energy
remains pretty constant in time. Since the "thermal energy" $E_{th}$ is strictly conserved in time
the total pseudo-energy, thermal plus potential, may indeed be paralleled to a true conserved
quantity (Hamiltonian) for most of the time evolution of the system, except a very short 
initial transient. 

\begin{figure}
 \includegraphics[scale=1.2]{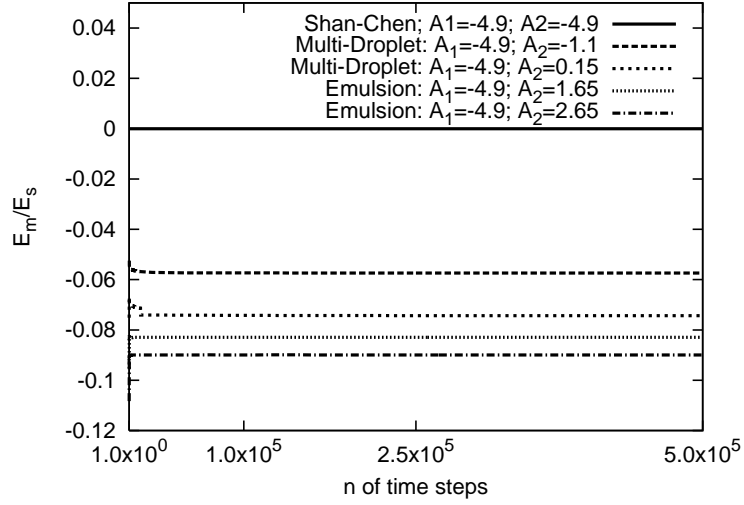}
\caption{\small{Time evolution of the ratio 
$E_m/E_s$ of the energy associated with the second to first-belt, Eq. (\ref{eq:e2}). 
The SC line (full) represents the ground state. As $A_2$ overcomes
$A_{2c}$, the energy level increases sensibly due to 
the interfacial contribution.
This figure shows the importance of repulsive mid-range 
interaction in the emulsion region.
This interaction is responsible for arresting the droplet coalescence sustained by short-range interaction, thereby promoting increased order
in the geometrical distribution of the droplets.}}
\label{Fig:7}
\end{figure}

In figure (\ref{Fig:7}), the time evolution of the ratio of first-belt to second-belt  pseudo-energy, for increasing values of the parameter $A_2$, is shown. 
Here again, after a very short transient, the ratio settles down
to a constant value, which is an increasing function of $A_2$.
To be noted that in all cases the ratio  is less than $10$ percent.
Yet, the effect on the surface/volume ratio of the fluid configuration is a very sizeable
one, as we shall discuss shortly.
The ratio between interfacial and bulk components can be estimated as
$E_{surf}/E_{bulk} \sim \frac{A \Delta x}{V} \frac{\Delta x}{\delta}$, $\delta$ being
the width of interface, $A$ the interfacial area and $V=L^2$ the volume of the simulation box. 
We have checked that the total volume of the liquid phase is the same as in the
standard SC case, whereas the interfacial area grows roughly with the scaling relation 
area/volume $\sim n^{1/2}$, $n$ being the number of droplets.
This is simply explained in term of mass conservation:
the volume of a single droplet is given by $\pi R^2$, while with $n$ droplets, the same volume is given by 
$\pi n R_n^2$, so that  $R_n\approx R/\sqrt{n}$.
This argument together with the dependence of the number of droplets on 
the mid-range force $G_2$ shown in fig. \ref{Fig:3} gives a relation between 
the final average domain size and the force $R\sim G_2^{1/2}$. 
This non-linear dependence is seen if fig \ref{Fig:radius}.

The consistency between theoretical estimation and simulation results
has been checked. For instance for the case $A=0.15$.
Typical values 
are $A/V \sim 0.1$ and $\delta/\Delta x \sim 10$,
such that the surface energy should be of the order of $1\%$.
This is in line with the actual surface energy, as shown in figure \ref{Fig:5}.

\begin{figure}
 \includegraphics[scale=0.9]{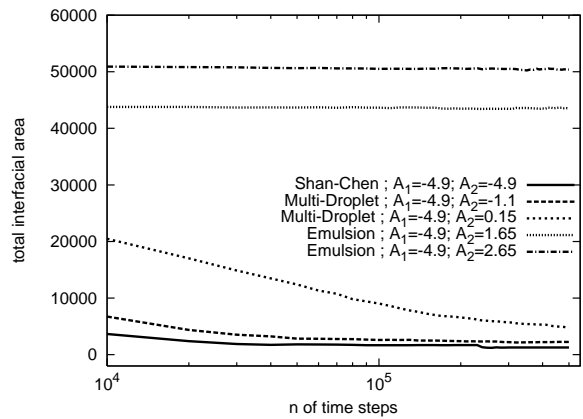}
\caption{\small{Total interfacial area, perimeter in 2D, as a function of time for different values of $A_2$.
This observable is used to monitor the onset of the transition between the multi-droplet region and the emulsion one.
In the multi-droplet region $A_2 \le A_{cr}$, the asymptotic limit is always the equilibrium state with one single droplet, that is the stable minimum in surface and, thus, in free-energy. It is clear, however, that the configuration obtained by changing $A_2$ has a different, and yet always finite, relaxation time. 
For $A_2 > A_{cr}$, the area of the liquid phase remains nearly constant 
in time and the relaxation time presents a sharp 
jump, virtually to an infinite value, thus signalling a 
phase-transition. 
The emulsion state is still metastable, but with a life-time much longer than the simulation-time.}}

\label{Fig:6}
\end{figure}

In figure \ref{Fig:6} we show the surface (perimeter in two-dimensions) 
of the multiphase fluid as a function of time for different values of $A_2$.
This is seen to go from roughly $5 \times 10^{-3}$ of the volume for the SC configuration, 
up to $0.19$ of the volume for the
emulsion-like configuration, thus showing a factor $40$ boost in surface/volume ratio, even though
the ``potential'' energy  in the second-shell is just a $10$ percent of the energy in the first shell,
as shown in fig. \ref{Fig:7}.
Such a dramatic boost shows that indeed a tiny amount of mid-range repulsion can cause
dramatic effects on the macroscopic fluid configuration.
From this time evolution, it is possible to extract a rough
estimate of the equilibrium relaxation-time of the system, 
namely the time necessary to relax to the minimum ``free-energy state'' (single-droplet). 
For the standard Shan-Chen model, this time has been measured to 
be $t_{sc}\approx10^3$.

\section{Conclusions}

Summarzing, the effects of mid-range repulsion in Lattice Boltzmann models  of single-component, non-ideal fluids are investigated. 
The simulations show that mid-range repulsive interactions promote the formation of  
spray-like, multi-droplet configurations, with droplet size directly related 
to the strength of the repulsive interaction.
Our results indicate that a small amount of mid-range repulsion can dramatically increase the 
surface/volume ratio of the multiphase fluid.

The present approach should offer an useful tool for the 
computational modelling of complex flow phenomena, such as 
atomization, spray formation and micro-emulsions, break-up phenomena 
and possibly glassy-like systems as well~\cite{Poo_92}.

\section{Acknowledgements}

 Helpful discussions with R. Benzi, L. Biferale, F. Toschi and A. Cavagna are kindly acknowledged.
We thank A. Lamura for his precious help.
SC's work is partially funded by a EU Marie-Curie ERG grant.
This work makes use of results produced by the PI2S2 Project managed by the Consorzio COMETA, a project co-funded by the Italian Ministry of University and Research (MIUR) within the Piano Operativo Nazionale “Ricerca Scientifica, Sviluppo Tecnologico, Alta Formazione” (PON 2000-2006). More information is available at http://www.pi2s2.it and http://www.consorzio-cometa.it.

\end{document}